\documentclass[a4paper,fleqn]{cas-sc}

\usepackage[authoryear]{natbib}
\usepackage{caption}
\usepackage{multirow}
\usepackage{colortbl}
\usepackage{enumitem}
\usepackage{threeparttable}
\usepackage{tabularx}
\usepackage{array}
\usepackage{longtable}
\usepackage{flafter} 

\usepackage[commandnameprefix=always,final,markup=default,authormarkup=none]{changes}
\newdefinition{RQ}{RQ}

\begin{document}
\let\WriteBookmarks\relax
\def\floatpagepagefraction{1}
\def\textpagefraction{.001}
\shorttitle{}
% \shortauthors{ANONYMOUS AUTHOR(S)}
\shortauthors{K. Qian et al.}

% \title[mode = title]{Evaluate before you scaffold in GenAI-Era: }

\title[mode = title]{Towards Reliable Generative AI-Driven Scaffolding: Reducing Hallucinations and Enhancing Quality in Self-Regulated Learning Support}

% previous
% \title[mode = title]{Mitigating Hallucinations and Enhancing Quality: Towards Automated Evaluation of Generative AI-Enabled Scaffolds for Self-Regulated Learning}

% \author{ANONYMOUS AUTHOR(S)}

\author[1]{Keyang Qian}[type=editor,
                        orcid=0000-0001-7118-3121
                        ]

\ead{keyang.qian@monash.edu}

\affiliation[1]{organization={Centre for Learning Analytics, Faculty of Informaiton Technology, Monash University},
    city={Melbourne},
    state={VIC},
    postcode={3800},
    country={Australia}}

\author[2]{Shiqi Liu}
\ead{liushiqi100@gmail.com}

\affiliation[2]{organization={School of Educational Science, Hunan Normal University},
city={Changsha},
state={Hunan},
    country={PR China}}

\author[1]{Tongguang Li}
\ead{tongguang.li@monash.edu}

\author[1]{Mladen Rakovi\'{c}}
\ead{mladen.rakovic@monash.edu}

\author[1]{Xinyu Li}
\ead{xinyu.li1@monash.edu}

\author[1]{Rui Guan}
\ead{rui.guan@monash.edu}

\author[3]{Inge Molenaar}
\ead{inge.molenaar@ru.nl}

\affiliation[3]{organization={Behavioural Science Institute, Radboud University},
    country={the Netherlands}}

\author[1]{Sadia Nawaz}
\ead{sadia.nawaz@monash.edu}

\author[1]{Zachari Swiecki}
\ead{zach.swiecki@monash.edu}

\author[4]{Lixiang Yan}
\cormark[1]
\ead{lixiangyan@mail.tsinghua.edu.cn}
\affiliation[4]{organization={School of Education, Tsinghua University},
city={Beijing},
    country={PR China}}

\author[1]{Dragan Ga\v{s}evi\'{c}}[type=editor,
                        % auid=002,bioid=2,
                        % role=Co-ordinator,
                        % orcid=0000-0001-9265-1908
                        ]
\cormark[1]
\ead{dragan.gasevic@monash.edu}

\cortext[cor1]{Corresponding author}

\begin{abstract}
Generative Artificial Intelligence (GenAI) holds a potential to advance existing educational technologies with capabilities to automatically generate personalised scaffolds that support students' self-regulated learning (SRL). While advancements in large language models (LLMs) promise improvements in the adaptability and quality of educational technologies for SRL, there remain concerns about the hallucinations in content generated by LLMs, which can compromise both the learning experience and ethical standards. \chdeleted[]{Furthermore, the probabilistic nature of LLM outputs may lead to inconsistent quality of scaffold generation.} To address these challenges, we proposed GenAI-enabled approaches for evaluating personalised SRL scaffolds before they are presented to students, aiming for reducing hallucinations and improving the overall quality of LLM-generated personalised scaffolds. Specifically, two approaches are investigated. The first approach involved developing a multi-agent system approach for reliability evaluation to assess the extent to which LLM-generated scaffolds accurately target relevant SRL processes. The second approach utilised the "LLM-as-a-Judge" technique for quality evaluation that evaluates LLM-generated scaffolds for their helpfulness in supporting students. We constructed evaluation datasets, and compared our results with single-agent LLM systems and machine learning approach baselines. Our findings indicate that the reliability evaluation approach is highly effective and outperforms the baselines, showing almost perfect alignment with human experts' evaluations. Moreover, both proposed evaluation approaches can be harnessed to effectively reduce hallucinations. Additionally, we identified and discussed bias limitations of the "LLM-as-a-Judge" technique in evaluating LLM-generated scaffolds. We suggest incorporating these approaches into GenAI-powered personalised SRL scaffolding systems to mitigate hallucination issues and improve the overall scaffolding quality.% exceed word limit, delete ", such as positional bias, self-enhancement bias, biases from sequential API calls, and verbosity bias"
\end{abstract}

% \begin{highlights}

% \item GenAI can produce scaffolds for self-regulated learning with questionable quality
% \item We propose an approach to evaluate scaffold quality and detect hallucinations.
% \item Multi-agent LLM approach can reliably detect quality issues in GenAI scaffolds
% \item LLM-as-a-Judge approach can reduce hallucinations with biases and moderate success
% \item Significant biases in LLM-as-a-Judge highlight the need for fairness strategies

% \end{highlights}


\maketitle

\section{Introduction}
Successful learners are typically characterised by their ability to effectively self-regulate their own learning \citep{boekaerts1997self,schunk2023self}. Effective self-regulated learners actively set their learning goals and consciously use and adapt different learning approaches to accomplish those goals \citep{winne1996metacognitive}. However, the literature shows that many learners have underdeveloped self-regulated learning (SRL) skills \citep{kirby2007self,difrancesca2016comparison}. For example, several studies showed that learners typically select to use ineffective learning approaches \citep{bjork_introducing_2023} and are inaccurate in monitoring of their learning \citep{gutierrez2022calibrating}. To address this challenge, researchers have been increasingly emphasising the need to support learners' SRL by providing personalised scaffolds for SRL in a scalable way \citep{van2023introduction,li2023analytics,lim2024students}. Some studies have demonstrated how learning analytics (LA)-based systems can be used to construct and provide personalised scaffolds for SRL, which in turn have been found to be positively associated with improved learning outcomes \citep{milikic2018measuring,lahza2022effects,munshi2023analysing,li2023analytics}. However, delivering personalised scaffolds automatically \chadded{during the student task} requires the formulation of rules that map insights from LA to actionable suggestions for specific learning tasks \chadded{\citep{backhaus2017assessing,schumacher2021investigating,li2023analytics,liu2024integrating,gunawardena2024personalized,li2025turning}}. Designing such rule-based systems can be labour-intensive, as it often involves creating complex if-then rules that classify learner's conditions into different cases and generating feedback tailored to each scaffolding case \citep{flasinski2016introduction,lim2023effects}. Additionally, these systems are highly task-dependent, meaning that any changes to the learning task requirements, materials, or goals \citep{van2023design,lim2023effects} may necessitate redesign of the system, limiting its scalability \citep{lim2024students,gunawardena2024personalized}.

Recent advances in Generative Artificial Intelligence (GenAI), particularly large language models (LLMs) like ChatGPT, offer significant promise for personalised scaffolding by overcoming the limitations of traditional rule-based approaches \citep{rudolph2023chatgpt,lin2023chat, li2025turning}. LLMs are AI agents capable of understanding and responding to user questions and instructions written in natural language (referred to as \textit{prompts}) in a human-like manner \citep{gozalo2023survey}. Ideally, a powerful LLM can interpret a given prompt that may include prior knowledge of the SRL model and useful SRL strategies, contextual information about the task, scaffolding instructions, the student's learner profile and SRL actions, thereby providing personalised scaffolds to the student. The functionality, adaptability, and cost-effectiveness of this approach are likely to further improve through rapid advancements in GenAI techniques \citep{inworld2023}. Consequently, the use of GenAI for personalised and scalable generation of SRL scaffolds is a promising research avenue.

Whilst GenAI techniques hold great promise to support SRL, researchers have also raised concerns about hallucinations that can often be found in LLM-generated contents\citep{yan2024promises}. Hallucinations in this context refer to a notable phenomenon that LLMs sometimes generate content that is irrelevant, made-up, or inconsistent \citep{lakera2024}. There is a heated debate among the GenAI research community about whether hallucinations are an inevitable aspect of LLM generation \citep{sangare2023}. One prevalent perspective \citep{xu2024hallucination, banerjee2024llms} suggested that hallucinations are indeed unavoidable for LLMs. We note that scaffolding students with LLM-generated content that contains hallucination may break the safety, reliability, and trustworthiness requirements of AI-powered learning technologies \citep{alfredo2024human}. However, limited studies have considered how to mitigate this issue with regard to GenAI-powered SRL scaffolding. Specifically, LLMs generate text in a randomised way depending on statistical probability instead of trying to produce semantically meaningful content as done by human educators \citep{zhao2023survey,dentella2024testing}. LLM generation may pose the risk of producing content that is not always relevant or meaningful even when there is no hallucination \citep{zhao2023survey}. This may make the quality of LLM-generated scaffolds, which are generated automatically in real-time without validation, less stable than traditional rule-based scaffolds, which are thoughtfully pre-designed by human educators \citep{li2023analytics}. It is often a labour-consuming job for educators to manually evaluate those LLM-generated scaffolds before they are sent to students, given that each scaffold may be different from other scaffolds due to the personalised nature of this approach and the number of students can be large indicating a need for scalable approaches. Consequently, there is a need to develop robust methodological approaches that can automatically evaluate the quality of LLM-generated scaffolds before providing them to students. Such methods can be used to automatically screen out low-quality and hallucinated LLM-generated scaffolds, triggering LLM regeneration of the scaffolds, and select high-quality scaffolds from multiple LLM-generated scaffolds, before providing scaffolds to students. This process aims to reduce the workload of educators while enhancing the educational effectiveness of LLM-generated scaffolds.

In this paper, we present the findings of a study aimed at improving the quality of LLM-generated scaffolds for SRL by addressing existing research gaps and investigating GenAI-enabled approaches for evaluating personalised scaffolds for SRL before they are presented to students. Specifically, we contribute two LLM-based approaches for evaluating scaffold reliability and quality. The first approach uses LLMs to assess the reliability of LLM-generated scaffolds based on how well they target the SRL processes \chadded{they need to support as instructed in the scaffold for SRL generation prompts}\chdeleted{outlined in the theoretical model of SRL proposed by winne1998studying}. The second approach adopts the \textit{LLM-as-a-Judge} method \citep{zheng2024judging} to evaluate the quality of multiple scaffolds and select the most helpful one with feedback according to predefined criteria in the prompt. This method aims to choose an optimal scaffold aligned with educational goals and includes structured elaborations that can support further human-in-the-loop evaluation and approval if needed. We evaluated 
\chreplaced{and showed the effectiveness of both approaches}{both approaches for their effectiveness} in identifying hallucinations \chdeleted[]{compared to human experts , and noted potential biases of LLM-as-a-Judge, such as positional bias, self-enhancement bias, biases from sequential API calls, and verbosity bias zheng2024judging,chen2024humans}. The findings contribute to the growing body of research on leveraging GenAI in education by presenting practical methodologies for improving the trustworthiness, reliability, and quality of LLM-generated scaffolds. \chreplaced{Another contribution is that}{Furthermore,} our work \chadded{is the first to} highlight \chadded{and  empirically test} potential bias issues in LLM-as-a-Judge evaluation \chreplaced{of LLM-generated educational feedback, which include position bias, self-enhancement bias, sequential API calling bias and verbosity bias,}{processes} and provides insights into mitigating these biases, paving the way for \chreplaced{applying LLM-as-a-Judge method in educational feedback evaluation more reliably.}{more equitable and effective applications of GenAI in supporting self-regulated learning at scale.} 

\section{Background}

\subsection{Self-Regulated Learning and COPES}
\label{sec:COPES}
SRL plays a pivotal role in students’ active engagement in learning by “monitoring, controlling, and regulating aspects of their cognition, motivation, behaviour, and context to meet task demands and build upon prior knowledge” \citep[p. 119]{greene2013two}. Students' academic performance and SRL processes are closely intertwined, throughout various stages of education \citep{dignath2008,vosniadou2020a,kesuma2021}. Specific SRL strategies, such as goal setting, self-monitoring, and strategic planning, have been observed to have a positive impact on academic performance and achievement \citep{zimmerman2011,rakovic2022using,Panadero2017,qin2019english,schunk2017,karlen2017}. In the theoretical model of SRL formulated by Winne and Hadwin \citep{winne1996metacognitive,winne1998studying,alvarez2022tools}, SRL consists of four phases: task definition, goal setting and planning, enactment of learning tactics and strategies, and adaptation. At the core of these phases is COPES that offers a detailed depiction of this learning process. COPES stands for \textit{Conditions}, \textit{Operations}, \textit{Products}, \textit{Evaluations}, and \textit{Standards} \citep{winne2023roles,winne1998studying}. These elements interlink dynamically to form the task-level structure that underscores the significance of SRL. \textit{Conditions} encompass external and internal factors like available resources or personal anxiety. \textit{Operations} detail the processes used to grasp information and consolidate knowledge. \textit{Products} refer to the outcomes attained. \textit{Evaluations} involve gauging these outcomes against set Standards, leading to informed decisions about subsequent strategies \citep{winne2017cognition}. Thus, COPES captures the comprehensive, cyclical nature of SRL, reinforcing the importance of monitoring and adjusting learning strategies to achieve educational goals.

\subsection{Scaffolding in Self-Regulated Learning}

Scaffolding is an educational construct defined as structured guidance provided to learners to foster skill acquisition within a specific learning context \citep{pea2018social}.This guidance is increasingly integrated into digital learning environments through technological scaffolding interventions \citep{siadaty2016measuring,siadaty2016associations}, often in the form of feedback messages. Automatic scaffolding systems, a key approach to providing such support \citep{lehtinen1999computer,wang2022design}, are the focus of this work. In promoting SRL, scaffolding seeks to empower students to effectively self-regulate their learning by supporting strategies such as goal setting, self-monitoring, and strategic planning \citep{zimmerman2011,rakovic2022using,Panadero2017,qin2019english,schunk2017,karlen2017}. Given the complexity of SRL processes and the frequent challenges students face in deploying effective learning strategies independently \citep{guo2022using,jovanovic2017learning,greene2000qualitative,bannert2009promoting}, scaffolding becomes essential. It serves to bolster learners' capacity to navigate and master the intricacies of SRL, ultimately enhancing their academic outcomes \citep{nelson1990metamemory}. 

\subsection{Existing Rule-based Approaches for Personalised Scaffolding}

Research into computer-based automated \textit{fixed SRL scaffolding} suggests that uniform feedback for all students is linked to improvements in academic performance and learning processes across various contexts \citep{li2023analytics}, examined in relation to effectiveness, persistence, group dynamics, and different educational infrastructures \citep{bannert2015short,sonnenberg2019using,dignath2008,vosniadou2020a,kesuma2021}. Beyond the success of \textit{fixed SRL scaffolding}, researchers also seek to offer automated \textit{personalised scaffolding for SRL} that provide adaptive support to tailor different student needs \citep{azevedo2005}. To do this, various researchers have proposed the use of real-time analytics of traces about students' SRL to determine which process students are engaging and critical for a given scaffolding point, and provide scaffolds for these SRL processes to the students (\citep{lim2023effects,van2023design,lim2024students}. Several studies have suggested that personalised scaffolding approaches have the potential to outperform fixed scaffolding approaches \citep{wong2021examining,guo2022using,van2023design,pardo2019using}. 

Research before the GenAI era implemented automatic personalised scaffolding for SRL by employing complex rule-based algorithms that are specially designed for a specific learning task with specific learning contexts in a digital learning environment \citep{li2023analytics}. The algorithms would code different insights obtained through the use of real-time analytics and provide pre-designed fixed scaffolds to the students accordingly \citep{azevedo2005,van2023design,lim2023effects,li2023analytics,lim2024students}.

Despite their benefits, rule-based scaffolding approaches face limitations. Designing such approaches demands considerable effort because they rely on creating intricate if-then rules and corresponding feedback for each scaffolding case. These designs are highly task- and context-specific, meaning that any changes to the task or context necessitate redesigning the approach. This limitation hinders their scalability \citep{li2023analytics,lim2023effects,van2023design,lim2024students,gunawardena2024personalized}. Moreover, evaluating and improving the quality of numerous, personalised scaffolds provided as feedback messages timely remains an unsolved issue, as poorly designed feedback messages may hinder learners’ SRL processes \citep{shih2010development,munshi2023analysing,alvarez2022tools}.

\subsection{Generative AI in Scaffolding}

\label{bak:genai_in_scaffolding}

The rise of GenAI offers a promising solution to the limitations of rule-based scaffolding approaches by leveraging the capabilities of LLMs, such as GPT-4 \citep{bubeck2023sparks,lin2023chat,li2025turning}. Modern LLMs can generate personalised feedback messages as a form of personalised scaffolds that match individual student needs by comprehending and generating human language in real time \citep{rudolph2023chatgpt,lin2023chat}. Utilising intuitive prompt frameworks, LLMs hold the potential to efficiently incorporate task context, SRL concepts, and specific student requirements without extensive pre-programming, thus facilitating the scalability of personalised SRL scaffolding systems \citep{kikalishvili2023unlocking,fariani2023systematic,schiff2021out,chang2023educational,liusocraticlm,qian2025scalefeedback}.

\label{bak:hallucinations}

Deploying GenAI in educational contexts poses challenges, particularly the risk of LLMs generating biased or factually inaccurate responses, known as hallucinations \citep{lakera2024,banerjee2024llms,sangare2023}. Existing research have shown that all current LLMs suffer from the hallucination problem, and much research considered hallucination as an inevitable nature of LLMs \citep{sangare2023,xu2024hallucination,banerjee2024llms}.

\label{bak:types-hallucination}

Broadly, types of hallucinations include \textit{Context-conflicting Hallucinations} (self-contradiction in the LLM response), \textit{Input-conflicting Hallucinations} (generation contradicts with prompt instructions), and \textit{Fact-conflicting Hallucinations} (contradiction with facts) \citep{zhang2023siren,rawte2023survey}. Based on these pre-defined hallucination types, in the context of SRL scaffolding, Input-conflicting Hallucinations are the major type of hallucination we focused on, as the LLM generation of SRL scaffolds can: \textit{case a)} not strictly follow instructions given in the prompt of providing targeted support about SRL \citep{hattie2007power}; \textit{case b)} violate the required scaffold format instructed in the prompt \citep{zhang2023siren}; \textit{case c)} conflict with information (facts) given in the prompt \citep{long2024llms}; and \textit{case d)} violate other instructions in the prompt. Hence, addressing this hallucination type is crucial to ensuring the reliability, safety, and trustworthiness in personalised scaffold generated based on LLMs and thus follow the principles of human-centred LA/AIED systems \citep{alfredo2024human}. Nevertheless, there are currently no methods that can effectively evaluating scaffolds generated by LLMs prior to providing such automatically generated scaffolds, in order to mitigate hallucination issues in an educational context.

\label{bak:srl-process-parser}

To fill this research gap in automated mitigation of hallucinations in scaffolds generated with LLMs, we first aimed to assess the \textit{reliability} of LLM-generated scaffolds in targeting specific SRL processes. This can be achieved by evaluating how well these scaffolds align with the intended SRL processes outlined in the prompt, as described in \textit{case a)} in the prior paragraph. LLMs have shown remarkable capacity for such reliability evaluation tasks, achieving state-of-the-art performance across a wide range of natural language processing (NLP) tasks, such as text classification, even in zero-shot or few-shot settings, which means the prompt includes zero or only a few examples of how to perform the task \citep{qian2025dean,yang2024harnessing,soltan2022alexatm,xu2024x,de2024benchmarking,zhang2024pushing}. Leveraging this capacity, we utilised LLMs to assess scaffold reliability and address our first research question:

\begin{RQ}
To what extent can LLMs evaluate the reliability of LLM-generated scaffolds in targeting relevant SRL processes?
\end{RQ}

\label{bak:llm-structure}

In the context of developing a GenAI system for scaffold evaluation, there are two common LLM structures \citep{wu2023autogen}: the single-agent structure, where a single LLM agent is responsible for all tasks, and the multi-agent structure, where multiple LLM agents collaborate to complete tasks. The single-agent structure is simpler to implement and more cost-effective, while the multi-agent structure has the potential to achieve higher performance by leveraging task specialisation \citep{talebirad2023multi,han2024llm}. To address \textbf{RQ1}, we investigated and compared both structures to determine their effectiveness for scaffold reliability evaluation.

\label{bak:llm-as-judge}

In addition to reliability evaluation, LLMs can also perform \textit{quality} evaluation using a method known as \textit{LLM-as-a-Judge} \citep{zheng2024judging,huang2024empirical,weyssow2024codeultrafeedback}. This method enables an LLM to act as a Judge agent (i.e. a \textit{LLM Judge}) to compare two responses (e.g., scaffolds) to the same prompt based on predefined quality aspects, imitating human judgment preferences, reducing the need of human labellers in a scalable way. In this study, we employed the \textit{pairwise comparison} type of LLM Judge agent that has been widely discussed and tested \citep{zheng2024judging,shi2024judging,chen2024humans}. \cite{zheng2024judging} proposed a general prompting framework for this type of Judge that aims to evaluate and choose a better answer from two answers of a open-ended question, according to critical quality aspects specialised in the prompt, like compliance of prompts, readability, helpfulness, level of detail, depth and relevance. By using this prompt framework, high-performing LLM Judges like GPT-4 align closely with human judgment preferences, achieving over 80\% agreement—comparable to inter-human agreement levels, and it has also been applied in evaluating educational feedback \citep{zheng2024judging,koutcheme2024open,huang2024empirical,weyssow2024codeultrafeedback,zhang2023evaluating,liusocraticlm}, highlighting its potential of offering general feedback evaluation with human-level quality. Since one of the critical challenges of LLM-generated scaffolds is their variability in quality and personalisation levels due to the probabilistic nature of text generation with LLMs \chadded{(i.e., even with the same input, an LLM might produce slightly different or inconsistent outputs, leading to variations in the quality of generated scaffolds)} \citep{zhao2023survey,dentella2024testing}, the LLM-as-a-Judge approach is a promising approach to provide a scalable solution to this issue by automatically evaluating and selecting the highest-quality scaffold before delivery to students. Moreover, LLM Judges have shown promise in detecting hallucinations in LLM-generated content \citep{zhang2023language}. By systematically identifying and selecting scaffolds with fewer hallucination issues, this approach has the potential to mitigate the hallucination risks in LLM-based scaffolding systems. However, there has been limited research on the effectiveness of LLM-as-a-Judge approaches in evaluating the quality of LLM-generated scaffolds for SRL. To address this gap, we explored the applicability of LLM Judges in scaffold evaluation and defined the following research question:

\begin{RQ}
To what extent can the LLM-as-a-Judge approach improve the quality of LLM-generated scaffolds by mitigating hallucination issues?
\end{RQ}

\label{bak:bias-llm-as-judge}

While promising, the LLM-as-a-Judge approach may exhibit notable biases. For scaffold evaluation in particular, three bias issues previously identified \chadded{in the evaluations of LLM-as-a-Judges} by \cite{zheng2024judging,chen2024humans} are of concern: 1) \textbf{Position bias}, where LLMs show a preference for certain positional arrangements of responses in the prompts; 2) \textbf{Verbosity bias}, where LLMs favour longer responses even when they may be less relevant or effective; and 3) \textbf{Self-enhancement bias}, where an LLM Judge may favour content generated by itself or similar models—for instance, GPT-4 may assign a 10\% higher win rate to responses it generated as compared to those from other models \citep{zheng2024judging}. A common strategy for mitigating position bias is \textbf{swapping positions}, where the order of responses is reversed during evaluation, and a response is only considered the winner if it is preferred in both cases \citep{zheng2024judging}. However, the extent to which these or other biases manifest in LLM Judge evaluations of scaffolds for SRL has not yet been explored. To address this, we defined the following research question:

\begin{RQ}
To what extent are bias issues present in LLM-as-a-Judge evaluations of LLM-generated scaffolds?
\end{RQ}

By answering these three research questions, the current study aimed to advance the development of reliable and high-quality LLM-generated scaffolding systems for SRL. We propose two LLM-based scaffold evaluation approaches -- reliability evaluation and quality evaluation -- both of which ensure scaffold assessment occurs before scaffold delivery to students, addressing the instable generation quality and hallucination challenges in LLM-generated scaffolds. Additionally, the investigation into bias mitigation strategies further contributes to improving the trustworthiness and fairness of LLM-based scaffold evaluation methods.

\section{Methods}

\begin{figure}[!ht]
  \centering
  \includegraphics[width=\textwidth]{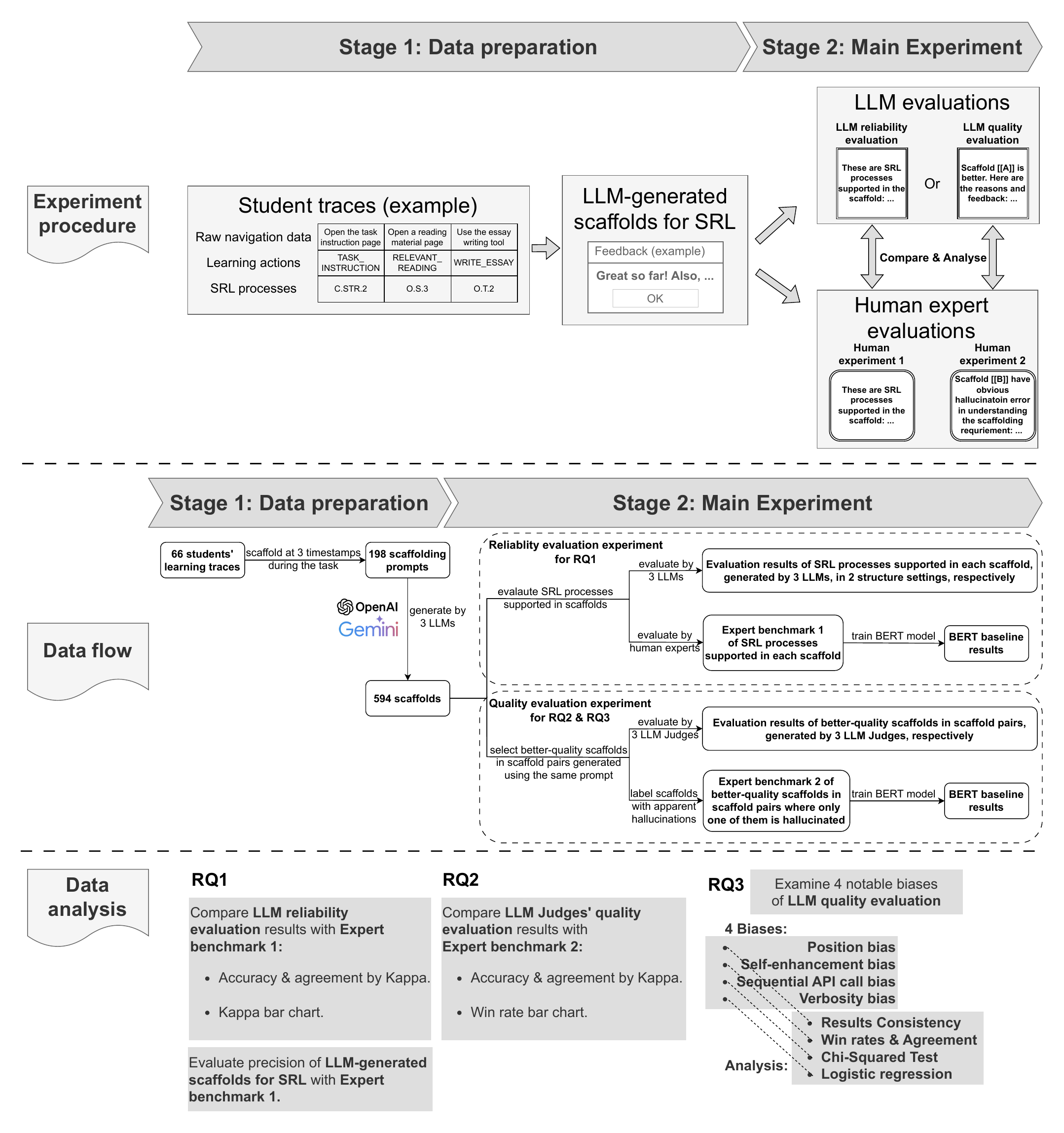}
  \caption{Study procedures and data analysis of the two proposed LLM-based evaluation approaches. \chdeleted[]{The upper half of the figure shows the experiment procedure from left to right, and the lower half of the figure illustrates the data analysis content for each research question.}}
  \label{fig:exp-process}
\end{figure}

\chreplaced{To evaluate LLM-generated scaffolds before they're provided to students,}{To systematically evaluate LLM-generated scaffolds for personalised SRL support,}  we conducted a study employing two novel GenAI-enabled automated evaluation approaches: \textit{reliability evaluation} and \textit{quality evaluation}. These approaches were tested on a dataset of LLM-generated scaffolds designed for a secondary school reading-writing task context. The procedure is illustrated in Fig.~\ref{fig:exp-process}. First, the scaffold dataset was generated using the GenAI-enabled SRL scaffolding method. Next, the generated scaffolds were evaluated using the \textit{reliability evaluation} and \textit{quality evaluation} approaches. Results from these evaluations were compared against human expert annotations to assess the effectiveness and biases of the proposed methods.

The proposed approaches are aimed to be used to evaluate LLM-generated scaffolds before they are provided to students, so that this step can automatically screen out low-quality and hallucinated scaffolds, triggering LLM regeneration of the scaffolds, and choose high-quality scaffolds from multiple LLM-generated scaffolds.% Details of the idea of how the proposed LLM-generated SRL scaffold evaluation approaches can be used in GenAI-enabled automated scaffolding systems to improve scaffolding quality and mitigate hallucinations are discussed in Appendix \ref{appx:workflow}.

\subsection{LLM-generated Scaffold Dataset}
\label{sec:scaffold-generation}

\chadded{We collected trace data by asking learners to do a multi-source reading-writing task. Participants were asked to read two sets of documents about two topics: (1) Artificial Intelligence and (2) Artificial Intelligence in Medicine. They were required to compose an essay in 200-300 words elaborating their vision of the application of AI in medicine in the future. The time limit for reading and writing the essay was 45 minutes. We enrolled 66 students (28 female, 37 male, and 1 non-binary) between the ages of 12 and 15 (M = 13.44, SD = 0.84) from two secondary schools in Australia, with 33 participants from each school. All participants were native English speakers. Informed consent was obtained from students, their parents, and teachers, and ethical clearance was granted by an Australian University \chadded{under human ethics application number 35965}.}

\chdeleted[]{The learning trace dataset used in the current study was collected within the context of a 45-minute multi-source reading-writing task for secondary school students, where the students (N=66) were asked to read multiple documents and compose an essay (see supplementary material Appendix A for the details about the study for data collection).}

\chadded{Using the trace data collected, we then employed a GenAI-enabled scaffolding method to simulate how LLMs could generate scaffolds for SRL to support secondary school students during their learning task, targeting critical SRL processes that are insufficiently done by students according to their traces. Other details about data collection are disclosed in supplementary material Appendix A.}

\chreplaced{W}{To address the three research questions, w}e first applied GenAI to generate a personalised scaffold dataset where the scaffolds target at learners’ distinct needs during the task based on the analysis of trace data about the use of SRL processes. The LLM-generated scaffold dataset was created using a GenAI-enabled SRL scaffolding method based on an SRL trace dataset (details of trace dataset collection are provided in \chdeleted[]{supplementary material} supplementary material Appendix A). The scaffolding method aimed to support specific SRL processes parsed from student SRL traces according to the COPES model \citep{winne1998studying,winne2023roles}. Specifically, we used a \chadded{SRL} processes library (see \chreplaced{see Table~\ref{tab:srl_lib}}{supplementary material Appendix B}) that aimed to addressed 15 SRL processes that were derived from the COPES model and that could be explicitly scaffolded. This process library was constructed following the same principles proposed by \cite{fan2022towards}.

\begin{table}[htp]
\centering
\caption{The process library for detection of SRL processes from \chadded{learner trace} action patterns.}
\label{tab:srl_lib}
\begin{tabular}{lllll}
\hline
\textbf{SRL Facets}  & \textbf{Constructs} & \textbf{Acronyms} & \textbf{Definition} & \textbf{Action Patterns} \\ \hline & & C.SAR.1        & & Table\_Of\_Content                             \\ \cline{3-3} \cline{5-5} 
                              & \multirow{-2}{*}{\begin{tabular}[c]{@{}l@{}}Surveying available\\ resources\end{tabular}}         & C.SAR.2                 & \multirow{-2}{*}{\begin{tabular}[c]{@{}l@{}}Learner develops perception of resources for the\\ task (e.g., tools for text annotation).\end{tabular}} & Try\_Out\_Tools \\ \cline{2-5} 
                              & \begin{tabular}[c]{@{}l@{}}Surveying task\\ requirements\end{tabular}                             & C.STR.2                 & \begin{tabular}[c]{@{}l@{}}Learner develops perception about features of\\ the task.\end{tabular}  & \begin{tabular}[c]{@{}l@{}}Task\_Overview/\\ Task\_Requriement/\\ Learning\_Goal/\\ Rubric (first time)\end{tabular}   \\ \cline{2-5} 
                              & \begin{tabular}[c]{@{}l@{}}Monitoring for\\ time constraints\end{tabular}                         & C.MTC.1                 & Learner oversees time left for the task.  & Timer      \\ \cline{2-5} 
\multirow{-5}{*}{Conditions}  & \begin{tabular}[c]{@{}l@{}}Monitoring for\\ task requirements\end{tabular}                        & C.MTR.2                 & \begin{tabular}[c]{@{}l@{}}Learner oversees state task requirements during\\ the learning session.\end{tabular}          & \begin{tabular}[c]{@{}l@{}}Task\_Overview/\\ Task\_Requriement/\\ Learning\_Goal\\ (after the first time)\end{tabular} \\ \hline
                              &  & O.S.1                   &  & Search\_Annotation     \\ \cline{3-3} \cline{5-5} 
                              & \multirow{-2}{*}{Searching}            & O.S.3                   & \multirow{-2}{*}{\begin{tabular}[c]{@{}l@{}}Surveying available sources and information and\\ comparing search entries to the standards.\end{tabular}}  & Page\_Navigation  \\ \cline{2-5} 
                              & & O.M.1                   & & Label\_Annotation       \\ \cline{3-3} \cline{5-5} 
                              &  & O.M.2                   &  & Create\_Highlight       \\ \cline{3-3} \cline{5-5} 
                              & \multirow{-3}{*}{Monitoring} & O.M.3    & \multirow{-3}{*}{\begin{tabular}[c]{@{}l@{}}Evaluating the match of information to a\\ profile of standards (e.g., highlighted\\ phrase and categorical tag assigned to it).\end{tabular}}     & \begin{tabular}[c]{@{}l@{}}Read\_Annotation/\\ Delete\_Annotation\end{tabular}               \\ \cline{2-5} 
                              & Assembling           & O.A.3                   & \begin{tabular}[c]{@{}l@{}}Learner creates a meaningful composite of\\ two or more units of information.\end{tabular}                     & Pastetext\_Essay     \\ \cline{2-5} 
                              & Rehearsing            & O.R.2               & Learner creates a copy of information.   & Write\_Essay\_Rehearsing                      \\ \cline{2-5} 
                              &                     &                      &                      &                      \\
                              &                      & \multirow{-2}{*}{O.T.1} &              & \multirow{-2}{*}{Create\_Note\_Translating}                   \\ \cline{3-3} \cline{5-5} 
                              &                      &                      &                      & \\
\multirow{-11}{*}{Operations} & \multirow{-4}{*}{Translating}                                        & \multirow{-2}{*}{O.T.2} & \multirow{-4}{*}{\begin{tabular}[c]{@{}l@{}}Learner manipulates input information to output\\ information while preserving critical information\\ properties (e.g., original meaning is preserved\\ and new related information is inserted).\end{tabular}} & \multirow{-2}{*}{Write\_Essay\_Translating} \\ \hline
Standards                     & \begin{tabular}[c]{@{}l@{}}Adopting/applying\\ standard based on\\ task instructions\end{tabular} & S.ASBTS.2               & { \begin{tabular}[c]{@{}l@{}}Criteria against which products are monitored\\ (e.g., task instructions and scoring rubric).\end{tabular}}                               & Open\_Planner                                      \\ \hline
\end{tabular}

\end{table}

We generated scaffolds at the start of 15th, 22th and 29th minute of the task for all 66 students. To generate the scaffolds, we calculated and compared the differences in frequencies of these processes between the high-performance tertile student group and the low-performance tertile student group in 7-14 min, 14-21 min and 21-28 min of the task to conclude the essential processes in these periods and their effect sizes. Next, we calculated individual students' counts of these essential processes in these 7-min periods. Finally, we included all information in LLMs' prompt input to let LLMs generate 100-word paragraph-style scaffolds, one scaffold for each scaffolding time point and each student, making each LLM's scaffold dataset containing 66 $\times$ 3 = 198 scaffolds. Template and examples of the prompts and the generated scaffolds are detailed in supplementary material Appendix B. We deployed three popular LLMs to generate the scaffolds: GPT-4-Turbo (OpenAI's gpt-4-1106-preview API), GPT-3.5-Turbo (OpenAI's gpt-3.5-turbo-1106 API), and Gemini-Pro (Google's gemini-pro API). GPT-4-Turbo served as a representative of state-of-the-art and expensive LLM options, while GPT-3.5-Turbo and Gemini-Pro served as cost-effective alternatives. According to the widely-used Chatbot Arena ranking list \citep{lmarena2023}, the general model abilities of these LLMs are ranked as GPT-4-Turbo > Gemini-Pro > GPT-3.5-Turbo. The overall scaffold dataset consisted of 594 scaffolds, evenly distributed across the three LLMs (198 per LLM $\times$ 3). This scaffold dataset was subsequently used to evaluate and compare the outcomes of the proposed approaches.

\subsection{Reliability Evaluation Approach}

\label{sec:reliability-evaluation-approach}

\begin{figure}[!ht]
  \centering
  \includegraphics[width=\textwidth]{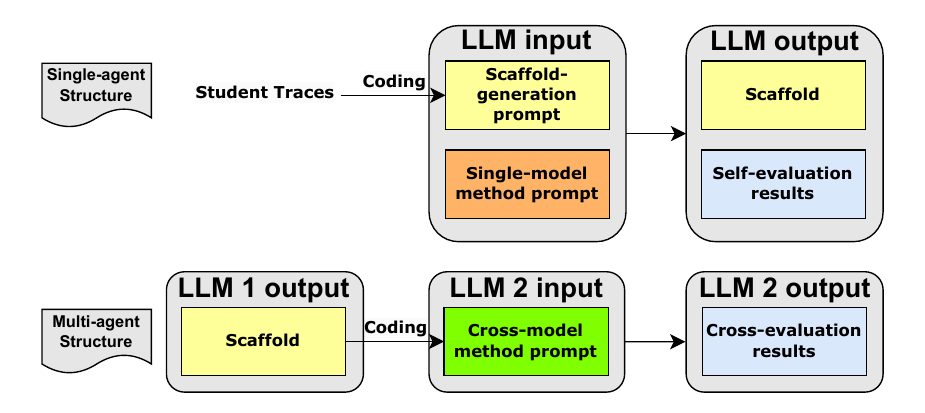}
  \caption{Comparison of single-agent structure and multi-agent structure. Yellow indicates scaffold generation components, orange represents single-agent structure components, green represents multi-agent structure components, and blue represents evaluation outputs which are in the same format for both configurations.}
  \label{fig:compare-single-and-cross}
\end{figure}

The reliability evaluation aimed to test how effectively LLM-generated scaffolds targeted the relevant SRL processes specified in the prompt\chadded{ – automatically before they are sent to students – in order to mitigate the risk that LLM-generated content may not always be relevant or meaningful \citep{zhao2023survey}}. This can be done with a \chadded{multi-label classification LLM} parser that inputs a scaffold and outputs the names of supported SRL processes in the scaffold in a structured manner. We compared the results of seven parsers, including six LLM-based parsers (three single-agent structure parsers and three multi-agent structure parsers) and one human expert parser benchmark. Fig.~\ref{fig:compare-single-and-cross} illustrates the key differences between the single-agent and multi-agent evaluation structures, showing how various components interact in each of these two methods. Fig. 10 in supplementary material Appendix C illustrates and compares the prompts of these two methods.

\subsubsection{Single-Agent Structure}  
In the single-agent structure, the same LLM that generated the scaffold was also used to report which SRL processes were supported. This was achieved by instructing the LLM to output a structured report of supported SRL processes as an additional paragraph following the generated scaffold. The LLM outputs were designed in a coded format that the data analysis programs can automatically read and analyse the processes results to facilitate automated analysis. Each single-agent parser labelled 198 scaffolds generated by its respective LLM, resulting in a total of 594 labelled scaffolds.

\subsubsection{Multi-Agent Structure}  
In the multi-agent structure, a separate LLM was assigned the task of reliability evaluation. This was implemented by constructing a prompt containing the scaffold, description of the SRL processes, and specific instructions for the evaluation task. The LLM used this prompt to parse the scaffold and report which SRL processes were supported, providing the output in a coded format same as the single-agent parser. Each multi-agent parser evaluated all 594 scaffolds in the dataset.

\subsubsection{Human Expert Benchmark 1}  

\label{sec:expert-benchmark-1}
To benchmark the LLM parsers, two expert coders with deep knowledge of the COPES model evaluated a subset of the scaffolds. Initially, the coders collaboratively labelled a training subset and drafted a detailed rubric for their evaluations. \chreplaced{Following this, the coders conducted a first-round inter-rater reliability test, achieving an almost-perfect level of overall agreement ($\kappa$ = 0.879, 95\% CI = [0.854, 0.903]). Subsequently, the coders reviewed and discussed the conflicting results, refining the labelling rubric based on their findings. During this process, two indistinguishable SRL processes (C.STR.2 and C.MTR.2) were merged into a single category, reducing the final number of SRL processes to 14. The coders then conducted a second-round inter-rater reliability test, labelling the SRL processes that had not reached an almost-perfect level of agreement in the first round. This round achieved an almost-perfect level of agreement for these processes as well \chadded{($\kappa$ = 0.9, 95\% CI = [0.819, 0.981])}. The labelling rubric is shared in Table 12 in Appendix F.1. Finally, the expert coders independently labelled 72 scaffolds evenly distributed among the three LLMs (24 per LLM) to minimise bias from the generator and serve as a benchmark for comparison.}{Inter-rater reliability was assessed in two rounds. achieving nearly perfect agreement in the second round (Cohen's Kappa > 0.8). During this process, two indistinguishable SRL processes (C.STR.2 and C.MTR.2) were merged into a single category, reducing the final number of SRL processes to 14. The expert coders independently labelled 72 scaffolds evenly distributed among the three LLMs (24 per LLM generator) to minimise bias from the generator \chadded{, and each 24 scaffolds per LLM generator were generated for the 3 scaffolding prompts of 8 random students (8 $\times$ 3 = 24) to preserve high variety and sample representative of supported processes instructed in the prompts}. And these final round labelling results served as a benchmark for comparison.}

\subsection{Quality Evaluation Approach}

\label{sec:quality-evaluation-approach}

The quality evaluation approach adopted the \textit{LLM-as-a-Judge} method proposed by \cite{zheng2024judging} to compare and select the best scaffolds according to critical quality aspects specialised in the prompt (the prompt is illustrated in Fig. 11 in supplementary material Appendix D). This method aimed to address the \chadded{unstable} quality and hallucination issues in LLM-generated scaffolds by selecting the most effective content for supporting SRL.

\begin{figure}[!ht]
  \centering
  \includegraphics[width=\textwidth]{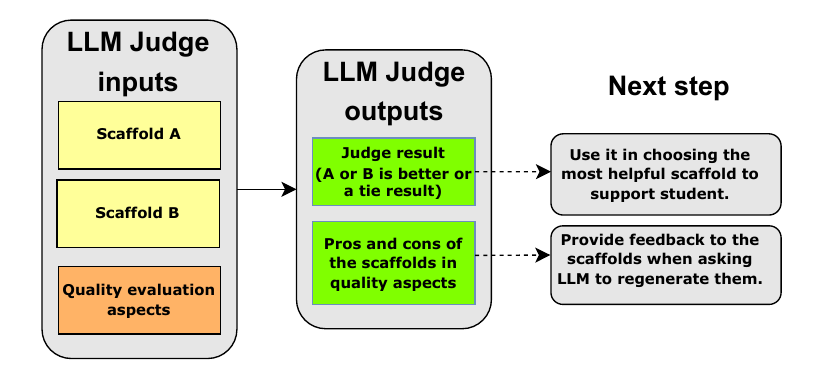}
  \caption{Illustration of the quality evaluation approach using LLM-as-a-Judge. Yellow highlights input scaffolds, orange displays quality evaluation requirements, and green represents the evaluation results. “Next step” shows where the Judge output components can be used in the GenAI-enabled scaffolding system.}
  \label{fig:qualitative-evaluation-illustration}
\end{figure}

\subsubsection{Evaluation Workflow}
\label{metd:evaluation-workflow}

Fig.~\ref{fig:qualitative-evaluation-illustration} illustrated the quality evaluation workflow. Three LLM models (GPT-4-Turbo, GPT-3.5-Turbo, and Gemini-Pro) acted as judges to compare and evaluate quality of pairs of scaffolds generated by the different LLMs using the same prompt inputs. As described in Sec.~\ref{sec:scaffold-generation}, each LLM produced 198 scaffolds in the same 198 prompt settings, resulting in 594 unique comparison pairs. For example, three LLM models used the same prompt to generate the scaffold that aimed to be sent to student A at the 15th minute of the task, hence each pair of the three generated scaffolds was compared by each LLM Judge. Moreover, as noted in previous research, the order in which one scaffold is mentioned before the other one in the prompt for a pair comparison may affect the judge’s decision \citep{zheng2024judging, chen2024humans}. To account for this, each pairwise comparison was repeated with the scaffold order swapped in the prompt (i.e., positions A and B in Fig.~\ref{fig:qualitative-evaluation-illustration}) to test for and eliminate potential position bias in the analysis.  This resulted in a total of 1,188 (i.e., 2 $\times$ 594) comparisons per LLM Judge. The LLM Judge assessed the scaffolds based on critical quality aspects specified in the evaluation prompt, including relevance, clarity, consistency, and alignment with the SRL processes. For each pair, the LLM Judge produced two outputs: i) a formatted result indicating the preferred scaffold, and ii) \chreplaced{justification of the choice,}{feedback} identifying the strengths and weaknesses of each scaffold with respect to the evaluation criteria. This \chreplaced{justification}{feedback} could subsequently be used to refine the prompts for scaffold regeneration if none of the scaffolds met the required quality\chadded{, which is not explored in this work but a future direction}.

\subsubsection{Human Expert Benchmark 2}  
\label{metd:benchmark2}

To assess the quality of LLM Judge results, human expert coders evaluated a subset of the scaffolds for hallucination errors. Coders identified scaffolds containing apparent hallucination errors that conflicted with the requirements stated in the original prompt (i.e., input-conflicting hallucinations as defined in Sec.~\ref{bak:types-hallucination}). \chreplaced{Coders first labelled some scaffolds together and drafted a detailed rubric. After that, coders conducted the first-round inter-rater test ($\kappa$=-0.09, n=81), discussed the conflicting results and refined the rubric. The labelling rubric is shared in Appendix F.2. Then, coders conducted the second-round inter-rater test which reached perfect agreement ($\kappa$=1) in the labelling (n=81).}{ Coders achieved almost perfect agreement in their final inter-rater reliability test (Cohen's Kappa > 0.8), where 162 scaffolds were used in the reliability tests.} The remaining 432 scaffolds, 144 per LLM generator, were then independently evaluated by human experts \chadded{in the final labelling round, which were then used to form a benchmark dataset for RQ2 that is discussed later in Sec.~\ref{sec:method-llm-judge}. Each of the 144 scaffolds per LLM generator were generated for the 3 scaffolding prompts of 48 random students (48 $\times$ 3 = 144) to preserve high variety and sample representative of hallucinations in the sample scaffolds.} A total of 48 out of these 432 scaffolds were labelled with apparent hallucination errors, where 29 of them were generated by GPT-3.5-Turbo and 19 of them were generated by Gemini-Pro.

\subsection{Analysis}
\subsubsection{RQ1: Reliability Evaluation}

To evaluate the reliability of LLM-generated scaffolds in supporting SRL processes, we compared the parsing performance of three LLM parsers (GPT-4-Turbo, GPT-3.5-Turbo, and Gemini-Pro) across two structural setups: \textit{single-agent structure} and \textit{multi-agent structure}, using expert human annotations as the benchmark. Human coders identified SRL processes supported in scaffolds for a subset of 72 scaffolds (24 per LLM scaffold generator, as mentioned in Sec.~\ref{sec:expert-benchmark-1}). For each LLM parser, we calculated accuracy ratios and Cohen's Kappa scores to measure alignment with the human expert coders' benchmark. The performance of multi-agent parsers was benchmarked for all scaffolds (n=594 parsed scaffolds), whereas single-agent parsers were evaluated based only on their respective generated scaffolds (n=198 parsed scaffolds per parser). In addition, we calculated the average single-agent structure performance across all three LLMs (n=594 scaffolds) to generalise performance. To further assess the targeted accuracy of scaffold generation, human annotations were used to categorize scaffolds into three groups: 1) \textit{Precise processes}---SRL processes explicitly instructed in the prompt and supported in the scaffold. 2) \textit{Error processes}---SRL processes supported in the scaffold but flagged as unnecessary support because students already mastered them, as per the prompt. 3) \textit{Irrelevant processes}---SRL processes supported in the scaffold but irrelevant to the prompt's context. We quantified precision, error rates, and irrelevance rates for each LLM scaffold generator (n=24 scaffolds per generator) to identify potential hallucination issues related to unsupported or irrelevant SRL processes.

\subsubsection{RQ2: Quality Evaluation}

\label{sec:method-llm-judge}
To assess the ability of LLM Judges for mitigating hallucinations in scaffolds, \chreplaced{we used human expert labelled results of whether scaffolds contain apparent hallucination errors or not to form a benchmark dataset. The final labelling round results of n=432 scaffolds were used, within which n = 48 hallucinated scaffolds were identified, as described in Sec.~\ref{metd:benchmark2}.}{we used human-labeled data where scaffolds with apparent hallucination errors (n = 48, as discussed in Sec.~\ref{metd:benchmark2}) were identified.  where scaffolds with apparent hallucination errors (n = 48, as discussed in Sec.~\ref{metd:benchmark2}) were identified.} Then, each hallucinated scaffold was paired with a non-hallucinated scaffold generated from the same prompt. In one case (32 instances), when three scaffolds (say, A, B, and C) were generated from a prompt, 32 cases had one hallucinated scaffold (e.g., A) and two non-hallucinated scaffolds (B and C). Each hallucinated scaffold was paired with both non-hallucinated scaffolds, yielding 2 pairs per instance. This gives 32 $\times$ 2 = 64 pairs. In the other case (8 instances), two scaffolds (e.g., A and B) were hallucinated while one scaffold (C) was non-hallucinated. Here, each of the hallucinated scaffolds was paired with the non-hallucinated scaffold, yielding 2 pairs per instance. This gives 8 $\times$ 2 = 16 pairs. To sum up, this process resulted in a total of 64 + 18 = 80 scaffold pairs. And as previously discussed (in Sec.~\ref{sec:quality-evaluation-approach}), each pairwise comparison was repeated with the scaffold order swapped in the prompt, resulting in n = 80 $\times$ 2 = 160 Judge comparison settings to be used as benchmark to evaluate LLM Judge results.
% To assess the ability of LLM Judges to mitigate hallucinations in scaffolds, we used the human-labelled data where scaffolds containing apparent hallucination errors (n=48 as just discussed in Sec.~\ref{metd:benchmark2}) were identified. The hallucinated scaffolds were paired with non-hallucinated scaffolds generated using the same prompts, resulting in a dataset of 80 (32 $\times$ 2 + 8 $\times$ 2, which is explained later) scaffold pairs and 160 (80 $\times$ 2) Judge comparison settings to be used as benchmark to evaluate LLM Judge results. For instance, when three LLM models were given the same prompt to generate scaffolds, they produced three different scaffolds: A, B, and C. In 32 cases, one of these scaffolds (e.g., scaffold A) was labeled as having an apparent hallucination issue, while the other two (B and C) were not. To analyse this, we paired the hallucinated scaffold (A) with each of the non-hallucinated scaffolds (B and C), forming two scaffold pairs: A-B and A-C. As previously discussed (in Sec.~\ref{sec:quality-evaluation-approach}), each pairwise comparison was repeated with the scaffold order swapped in the prompt, resulting in four Judge comparison settings: A-B, B-A, A-C, and C-A. In the other 8 cases, 2 of these three scaffolds, e.g., A and B, were labelled with hallucination and scaffold C was not, then there were still 2 scaffold pairs (A-C and B-C) and 4 Judge comparison settings.

For each scaffold pair, LLM Judges (GPT-4-Turbo, GPT-3.5-Turbo, and Gemini-Pro) have evaluated which scaffold better supported the SRL requirements and meet the quality standards, as discussed in the previous Sec.~\ref{sec:quality-evaluation-approach}. Human expert labelled results served as the benchmark, where LLM Judges were expected to prefer non-hallucinated scaffolds over hallucinated scaffolds. Accuracy and Cohen's Kappa values were calculated for each LLM Judge to assess their effectiveness in rejecting hallucinated scaffolds. To control for positional bias in LLM Judge decisions, scaffold positions in the evaluation prompt were systematically swapped, generating 1,188 comparisons (two swaps per scaffold pair for three LLM Judges). The average accuracy and agreement with the human benchmark were used to measure the effectiveness of LLM Judges in detecting hallucinations.

\subsubsection{RQ3: Bias Evaluation}

Given known biases in LLM-based quality evaluations (Sec.~\ref{bak:bias-llm-as-judge}), this analysis tested four specific biases impacting LLM Judges: \textit{position bias}, \textit{self-enhancement bias}, \textit{sequential API call bias}, and \textit{verbosity bias}.

\textbf{Position Bias.} Position bias in our scenario refers to the potential tendency of LLM Judges to prefer scaffolds based on their position in the prompt, regardless of intrinsic quality. To assess positional bias, we computed and compared distribution of results of each Judge, the percentage of cases where a Judge gives consistent results when swapping the order of two scaffolds in the Judge prompt (i.e. \textit{consistency}), the percentage of cases when a Judge favors the scaffold lying in the first and second position of the prompt (i.e. \textit{biased toward first} and \textit{biased toward second}, respectively). Both original and swapped results were included (n=1,188 comparisons per LLM Judge, as discussed in Sec.~\ref{metd:evaluation-workflow}) to minimize the influence of scaffold quality disparities.

\textbf{Self-Enhancement Bias.} Self-enhancement bias in our scenario occurs when an LLM Judge disproportionately favors scaffolds generated by itself over those produced by other generators. We analysed self-enhancement bias by comparing win rates for scaffold generators (e.g., GPT-4-Turbo vs. GPT-3.5-Turbo) across three LLM Judges. This analysis controlled for positional bias by combining results from both original and swapped positions (n=396 per comparison pairs, e.g., GPT-4-Turbo vs. GPT-3.5-Turbo; 198 $\times$ 2, 196 as each LLM scaffold generator generates 196 scaffolds as previously mentioned, 2 for combining results from both original and swapped positions). Tie results were excluded for a fair comparison of win rates. To measure disagreement between LLM Judges, pairwise Cohen's Kappa values were calculated (n=1,188 per LLM Judge, as discussed in Sec.~\ref{metd:evaluation-workflow}).

\textbf{Sequential API Call Bias.} Sequential API call bias in our scenario assessed whether Judge decisions were influenced by the outcomes of previous API calls. To analyse how prior API call results influenced subsequent calls, we conducted Pearson's Chi-Squared Tests of Independence comparing Judge decisions at sequential timestamps. Cramér's V values measured the effect size of sequential dependencies (i.e. the \textit{R} value). As each time we generate 1,188 Judge results for each LLM Judges (as discussed in Sec.~\ref{metd:evaluation-workflow}), 1,187 out of these 1,188 results had a prior API call result, making the number of observations n=1,187 in the analysis.

\textbf{Verbosity Bias.}\label{sec:metd-verbosity} Verbosity bias in our scenario measured the extent to which LLM Judges preferred longer scaffolds that exceeded word limits. We investigated verbosity bias by examining LLM Judge preferences for scaffolds exceeding word limits. Logistic regression models (see Eq.~\ref{eq:logistic_regression}) were constructed to predict Judge preferences based on scaffold word limits and scaffold generator comparison settings (SGC). The analysis included observations where only one scaffold exceeded the word limit, including 648 out of the total 1,188 comparison settings per LLM Judge.% To balance number of  and when  resulting in n=1,296 satisfied comparisons per LLM Judge. 

\begin{equation}
\begin{aligned}
\text{{LLM Judge result}} = \beta_0 &+ \beta_1 \times \text{{condition [exceed-word-limit]}} + \beta_{2,3,4,5,6} \times \text{{condition [SGC settings]}}
\end{aligned}
\label{eq:logistic_regression}
\end{equation}

Each logistic regression model estimated coefficients ($\beta$ values) for the exceed-word-limit condition and five SGC configurations, with negative coefficients indicating a preference for exceeding word-limit scaffolds. Model robustness was validated using the proportional odds assumption, multicollinearity diagnostics, and residual independence checks. Coefficient values and 95\% confidence intervals (CIs) across models were compared to assess the severity of verbosity bias among the three LLM Judges.

\subsubsection{Baseline: Traditional Machine Learning Approach}

Before the emergence of ChatGPT, BERT is a commonly used \chdeleted[]{state-of-the-art} pre-trained machine learning model for NLP tasks \citep{wolf2020transformers}, and a \chdeleted[]{a} ubiquitous baseline in text classification experiments \chadded{of research in recent years} \citep{rogers2021primer}. To establish performance baselines \chdeleted[]{ for the evaluation approaches}, we fine-tuned the \chadded{default} pre-trained BERT model (\texttt{bert-base-uncased}) \chadded{of sequence classification tasks} using the transformer library~\citep{wolf2020transformers}. \chadded{We used the pre-trained BERT tokenizer and the default AdamW optimizer in transformers library to train the model. For both the RQ1 and RQ2 BERT models, they were trained on 64\% of the dataset, validated on 16\%, and tested on 20\% following common machine learning practice~\citep{prechelt2002early}. We trained these models on the training dataset and stopped the training until there's no further improvement on the validation dataset loss, which is referred to as early-stopping method \citep{prechelt2002early}. Six random seeds were used to improve robustness, and average performance metrics were reported for precision, accuracy, and agreement with human-labelled data.}

For RQ1, \chreplaced{we imported the Bert-For-Sequence-Classification model from transformers library, and fine-tuned it in our RQ1 human expert benchmark dataset (introduced in Sec.~\ref{sec:expert-benchmark-1}), taking scaffolds as the input to predict the supported SRL processes in the scaffolds, which is a multi-label text classification task. We set number of model output labels as 15, which is the number of SRL process categories in our dataset. We searched the probability thresholds of different output labels that can maximise F1-score on the validation dataset, which is referred as SCut method \citep{yang2001study,fan2007study}. The model results on the test dataset provided performance baselines for LLM-based approaches, which are used for comparison in the Results Sec.~\ref{sec:rq1result}.}{we trained the BERT model to classify scaffolds into 15 SRL process categories from our human-labelled dataset.}

For RQ2, \chreplaced{we imported the Bert-For-Sequence-Classification model from transformers library, and fine-tuned it in our RQ2 human expert benchmark dataset (introduced in Sec.~\ref{metd:benchmark2}), taking the scaffold-comparison pairs as the input to predict correct Judge results on the dataset, which is a multi-class text classification task. We set number of model output labels as 3 which is the number of Judge preference categories (prefer A, prefer B, and neural), and the model chose the highest probability label as the predicted class outcome. Compared with LLM Judges, the text inputs of the BERT model simply removed the "system" and "user question" parts of prompt inputs of LLM Judges (illustrated in Supplementary material Appendix D) because they are the same text input component for all data points. The model results on the test dataset provided performance baselines for LLM-based approaches, which are used for comparison in the Results Sec.~\ref{sec:rq2result}.}{scaffold pair comparisons were classified into three preference categories using a fine-tuned BERT sequence classification model.}

\section{Results}

\subsection{RQ1: Reliability Evaluation of LLM-generated Scaffolds}

\label{sec:rq1result}

The multi-agent structure showed significantly better performance in all LLMs compared to their single-agent counterparts (Table~\ref{tab:metrics-llm-parser}). Among the multi-agent parsers, GPT-4-Turbo demonstrated the highest reliability, achieving near-perfect agreement with human results ($\kappa = 0.808$, 95\% CI [0.765, 0.852]; accuracy = 0.931, 95\% CI [0.915, 0.946]), followed by Gemini-Pro, which reached a substantial agreement level ($\kappa = 0.603$, 95\% CI [0.544, 0.663]; accuracy = 0.854, 95\% CI [0.832, 0.876]). GPT-3.5-Turbo in the multi-agent structure achieved moderate agreement with human benchmarks ($\kappa = 0.526$, 95\% CI [0.467, 0.585]; accuracy = 0.803, 95\% CI [0.778, 0.827]). In comparison, single-agent structures demonstrated weaker performance overall. Specifically, the single-agent GPT-4-Turbo parser yielded  accuracy = 0.830 (95\% CI [0.790, 0.870], $\kappa=0.464$, 95\% CI [0.338, 0.591]), closely followed by GPT-3.5-Turbo (accuracy = 0.821, 95\% CI [0.780, 0.862], $\kappa=0.482$, 95\% CI [0.364, 0.601]) and Gemini-Pro (accuracy = 0.786, 95\% CI [0.742, 0.830], $\kappa=0.428$, 95\% CI [0.311, 0.545]). When averaged across all single-agent structures, agreement was moderate ($\kappa=0.475$, 95\% CI [0.408, 0.543], accuracy = 0.813, 95\% CI [0.788, 0.837]). Importantly, none of the LLM configurations exceeded the human expert benchmark ($\kappa=0.869$, 95\% CI [0.854, 0.903], accuracy=0.956, 95\% CI [0.930, 0.981]), which remained the gold standard.

\begin{table}[hbpt]
\centering
\caption{Metrics of LLMs as parsers of supported SRL processes in the scaffolds.}
\label{tab:metrics-llm-parser}
\begin{threeparttable}
\begin{tabular}{lllllll}
\hline
\multicolumn{1}{l}{\multirow{2}{*}{Metric}} & \multicolumn{3}{c}{\textbf{Accuracy}}                                 & \multicolumn{3}{c}{\textbf{Agreement by Kappa}}  \\
\multicolumn{1}{l}{}                  & Value          & CI Lower       & \multicolumn{1}{l}{CI Upper}       & Value          & CI Lower       & CI Upper       \\ \hline
\textbf{Multi-agent structure}            &                &                &                                     &                &                &                \\
\multicolumn{1}{l}{GPT-4-Turbo}             & \textbf{0.931} & \textbf{0.915} & \multicolumn{1}{l}{\textbf{0.946}} & \textbf{0.808} & \textbf{0.765} & \textbf{0.852} \\
\multicolumn{1}{l}{GPT-3.5-Turbo}     & 0.803          & 0.778          & \multicolumn{1}{l}{0.827}          & 0.526          & 0.467          & 0.585          \\
\multicolumn{1}{l}{Gemini-Pro}        & 0.854          & 0.832          & \multicolumn{1}{l}{0.876}          & 0.603          & 0.544          & 0.663          \\ \hline
\textbf{Single-agent structure}           &                &                &                                     &                &                &                \\
\multicolumn{1}{l}{GPT-4-Turbo}       & \textbf{0.830} & \textbf{0.790} & \multicolumn{1}{l}{\textbf{0.870}} & 0.464          & 0.338          & 0.591          \\
\multicolumn{1}{l}{GPT-3.5-Turbo}     & 0.821          & 0.780          & \multicolumn{1}{l}{0.862}          & \textbf{0.482} & \textbf{0.364} & \textbf{0.601} \\
\multicolumn{1}{l}{Gemini-Pro}        & 0.786          & 0.742          & \multicolumn{1}{l}{0.830}          & 0.428          & 0.311          & 0.545          \\
\multicolumn{1}{l}{Average}           & 0.813          & 0.788          & \multicolumn{1}{l}{0.837}          & 0.475          & 0.408          & 0.543          \\ \hline
\multicolumn{1}{l}{\textbf{Baseline}} &                &                & \multicolumn{1}{l}{}               &                &                &                \\
\multicolumn{1}{l}{BERT model}        & 0.613          & 0.541          & \multicolumn{1}{l}{0.685}          & -0.10          & -0.193         & -0.007         \\
\multicolumn{1}{l}{Human expert\chadded{s}\tnote{\textbf{\dag}}}     & 0.956          & 0.930          & \multicolumn{1}{l}{0.981}          & 0.869          & 0.854         & 0.903          \\ \hline
\end{tabular}
\begin{tablenotes}
        \footnotesize
        \item[\textbf{\dag}] 
        In-between consistency and agreement of two human experts in the first-round inter-rater test of labelling when they have already discussed and preliminarily settled a labelling rubric.
\end{tablenotes}
\end{threeparttable}
\end{table}

\begin{figure}[!ht]
  \centering
    \includegraphics[width=0.7\textwidth]{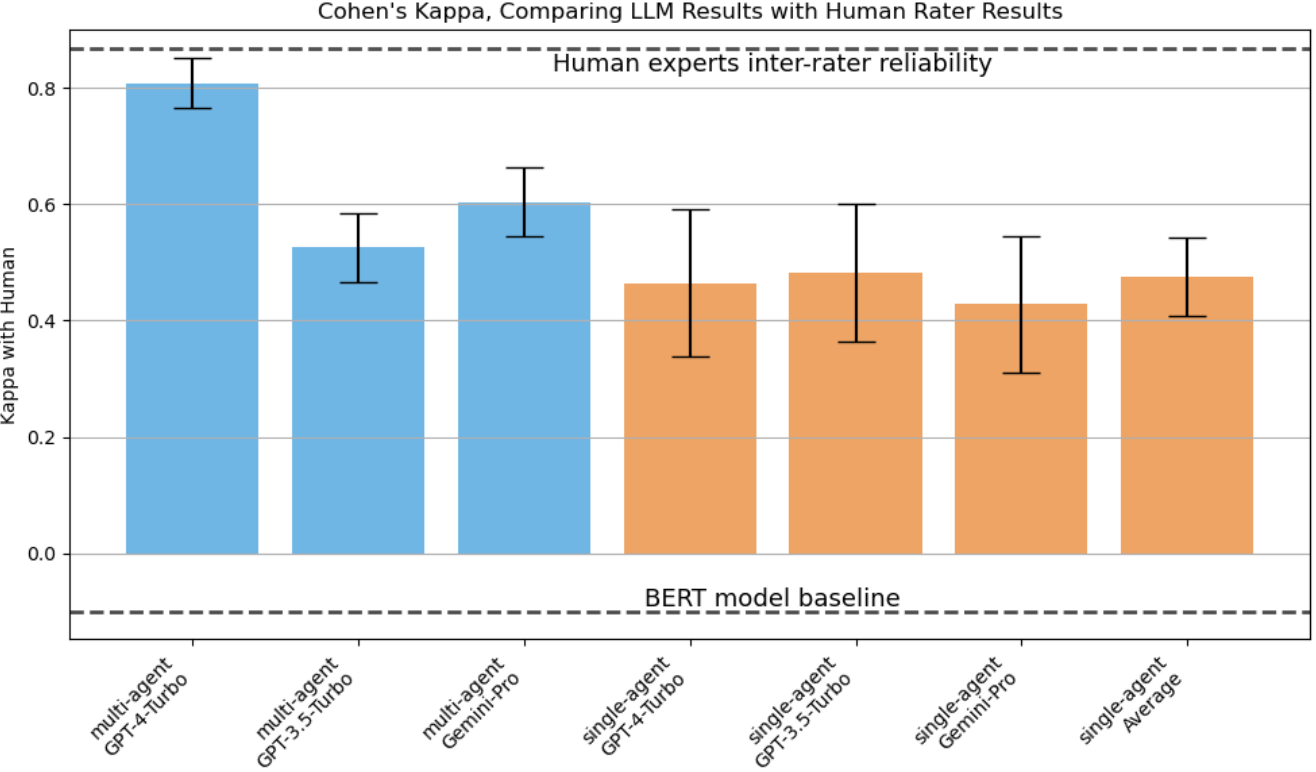}
      \caption{Bar chart of reliability of different LLM SRL process parser settings. Error bars represent 95\% CI.}
      \label{fig:parser_kappa_bar}
\end{figure}

The performance of LLM parsers also significantly outperformed the machine learning baseline represented by a fine-tuned BERT model ($\kappa=-0.100$, 95\% CI [-0.193, -0.007], accuracy = 0.613, 95\% CI [0.541, 0.685]). \chdeleted[]{This suggests that the LLM-based parsers provide a substantial improvement over traditional natural language processing approaches in parsing SRL processes.} Accuracy and agreement results further indicate that GPT-4-Turbo in the multi-agent structure consistently outperformed all other LLM configurations, aligning significantly closer to human benchmarks when compared to Gemini-Pro and GPT-3.5-Turbo. Fig.~\ref{fig:parser_kappa_bar} visually highlights these hierarchical differences, where the GPT-4-Turbo multi-agent parser achieved the highest accuracy ratios and agreement levels, followed by Gemini-Pro and then GPT-3.5-Turbo, while no substantial differences were observed among single-agent parsers within their respective CIs.

\label{sec:RQ1_result}

Metrics assessing the alignment of LLM-generated scaffold with the specific SRL processes instructed in the prompts are detailed in Table~\ref{tab:metrics-scaffold-target}. Scaffolds generated by GPT-4-Turbo demonstrated the highest precision (0.613, 95\% CI [0.503, 0.724]) and the lowest irrelevance rate (0.227, 95\% CI [0.132, 0.321]), indicating its ability to generate scaffolds more consistently aligned with the intended SRL processes. GPT-3.5-Turbo achieved moderate precision (0.528, 95\% CI [0.441, 0.614]) and irrelevance rate (0.283, 95\% CI [0.205, 0.362]), while Gemini-Pro demonstrated significantly lower levels of precision (0.355, 95\% CI [0.264, 0.446]) but the lowest error rate (0.093, 95\% CI [0.038, 0.149]). Despite this, all LLMs produced scaffolds with precision below 73\%, irrelevance rates above 13\%, and non-negligible error rates exceeding 3\%, reflecting ongoing limitations in the alignment between LLM scaffold generation and scaffold-generation requirements.

\begin{table}[!ht]
\caption{Metrics of to what extent scaffolds generated by different LLMs are targeted as instructed in the scaffold-generation prompt.}
\label{tab:metrics-scaffold-target}
\centering
\begin{tabular}{llllllllll}
\hline
\multirow{2}{*}{Metric} & \multicolumn{3}{c}{\textbf{Precision}}                                & \multicolumn{3}{c}{\textbf{Error Rate}}                               & \multicolumn{3}{l}{\textbf{Irrelevance Rate}}    \\
                        & Value          & CI Lower       & \multicolumn{1}{l}{CI Upper}       & Value          & CI Lower       & \multicolumn{1}{l}{CI Upper}       & Value          & CI Lower       & CI Upper       \\ \hline
GPT-4-Turbo             & \textbf{0.613} & \textbf{0.503} & \multicolumn{1}{l}{\textbf{0.724}} & 0.160 & 0.077 & \multicolumn{1}{l}{0.243} & \textbf{0.227} & \textbf{0.132} & \textbf{0.321} \\
GPT-3.5-Turbo           & 0.528          & 0.441          & \multicolumn{1}{l}{0.614}          & 0.189          & 0.121          & \multicolumn{1}{l}{0.257}          & 0.283          & 0.205          & 0.362          \\
Gemini-Pro              & 0.355          & 0.264          & \multicolumn{1}{l}{0.446}          & \textbf{0.093}          & \textbf{0.038}          & \multicolumn{1}{l}{\textbf{0.149}}          & 0.551          & 0.457          & 0.646          \\
Average                 & 0.489          & 0.433          & \multicolumn{1}{l}{0.544}          & 0.149          & 0.109          & \multicolumn{1}{l}{0.189}          & 0.362          & 0.309          & 0.416          \\ \hline
\end{tabular}
\end{table}

Overall, these findings suggest that LLM-based parsers, particularly in a multi-agent setup with GPT-4-Turbo, significantly outperformed traditional machine learning approaches and offered reliability close to human benchmarks. Despite this, we showed that scaffold generation by all tested LLMs suffered from notable inaccuracies, including hallucinations of irrelevant or erroneous SRL processes, underscoring the importance of the proposed reliability evaluation that can screen out those unreliable LLM-generated scaffolds before scaffolding to reduce these issues.

\subsection{RQ2: Mitigating Hallucinations by LLM Judges}

\label{sec:rq2result}

As shown in Table~\ref{tab:judge-hallucination}, among the three LLM Judges, GPT-4-Turbo demonstrated the highest performance with an accuracy of 0.70 (95\% CI [0.63, 0.77]) and a moderate level of agreement with human annotations ($\kappa = 0.40$, 95\% CI [0.28, 0.52]). The accuracy results for GPT-4-Turbo were significantly above the levels of both the weaker LLM Judges and the baseline BERT model (Accuracy = 0.49, 95\% CI [0.42, 0.57]; $\kappa = -0.01$, 95\% CI [-0.15, 0.13]). Furthermore, GPT-4-Turbo's accuracy surpassed the chance-level performance (coin toss; accuracy = 0.50), as the lower bound of its confidence interval was greater than 0.50. This indicates that GPT-4-Turbo could reliably select better scaffolds and significantly mitigate the issue of hallucination errors when acting as a Judge. However, even GPT-4-Turbo achieved only a moderate alignment with human-labelled benchmark, suggesting room for improvement in its ability to eliminate hallucinations entirely. In contrast, the other two LLM Judges, GPT-3.5-Turbo and Gemini-Pro, exhibited significantly lower performance, with accuracies of 0.48 (95\% CI [0.40, 0.56]) and 0.52 (95\% CI [0.44, 0.60]), respectively. The inclusion of chance-level accuracy (0.50) within the confidence intervals for both models implies that these LLM Judges were not reliably avoiding unreliable scaffolds and hallucination errors. In terms of agreement with the human expert labelled benchmark, GPT-3.5-Turbo displayed a poor level of agreement ($\kappa = -0.04$, 95\% CI [-0.20, 0.11]), while Gemini-Pro achieved a minimal level of agreement ($\kappa = 0.04$, 95\% CI [-0.11, 0.19]). These findings closely align with the poor performance of the BERT model baseline, further emphasising the limitations of using weaker language models for this quality evaluation task.

\begin{table}[!ht]
\caption{Metrics of to what extent do LLM Judges correctly not prefer human-labelled unreliable scaffolds with apparent hallucination errors of violating scaffolding requirement, when comparing with other scaffolds.}
\label{tab:judge-hallucination}
\begin{tabular}{lllllll}
\hline
\multicolumn{1}{l}{\multirow{2}{*}{Metric}} & \multicolumn{3}{c}{\textbf{Accuracy}}                              & \multicolumn{3}{c}{\textbf{Agreement by Kappa}}    \\
\multicolumn{1}{l}{}              & Value & CI Lower & \multicolumn{1}{l}{CI Upper} & Value & CI Lower & CI Upper \\ \hline
\textbf{LLM Judge}                 &       &          &                               &       &          &          \\
\multicolumn{1}{l}{GPT-4-Turbo}                      & \textbf{0.70} & \textbf{0.63} & \multicolumn{1}{l}{\textbf{0.77}} & \textbf{0.40} & \textbf{0.28} & \textbf{0.52} \\
\multicolumn{1}{l}{GPT-3.5-Turbo} & 0.48  & 0.40     & \multicolumn{1}{l}{0.56}     & -0.04 & -0.20    & 0.11     \\
\multicolumn{1}{l}{Gemini-Pro}    & 0.52  & 0.44     & \multicolumn{1}{l}{0.60}     & 0.04  & -0.11    & 0.19     \\
\multicolumn{1}{l}{Average}       & 0.57  & 0.52     & \multicolumn{1}{l}{0.61}     & 0.13  & 0.05     & 0.21     \\ \hline
\textbf{Baseline}                  &       &          &                               &       &          &          \\
\multicolumn{1}{l}{BERT model}    & 0.49  & 0.42     & \multicolumn{1}{c}{0.57}     & -0.01 & -0.15    & 0.13     \\ \hline
\end{tabular}
\end{table}

\begin{figure}[!ht]
  \centering
    \includegraphics[width=0.8\textwidth]{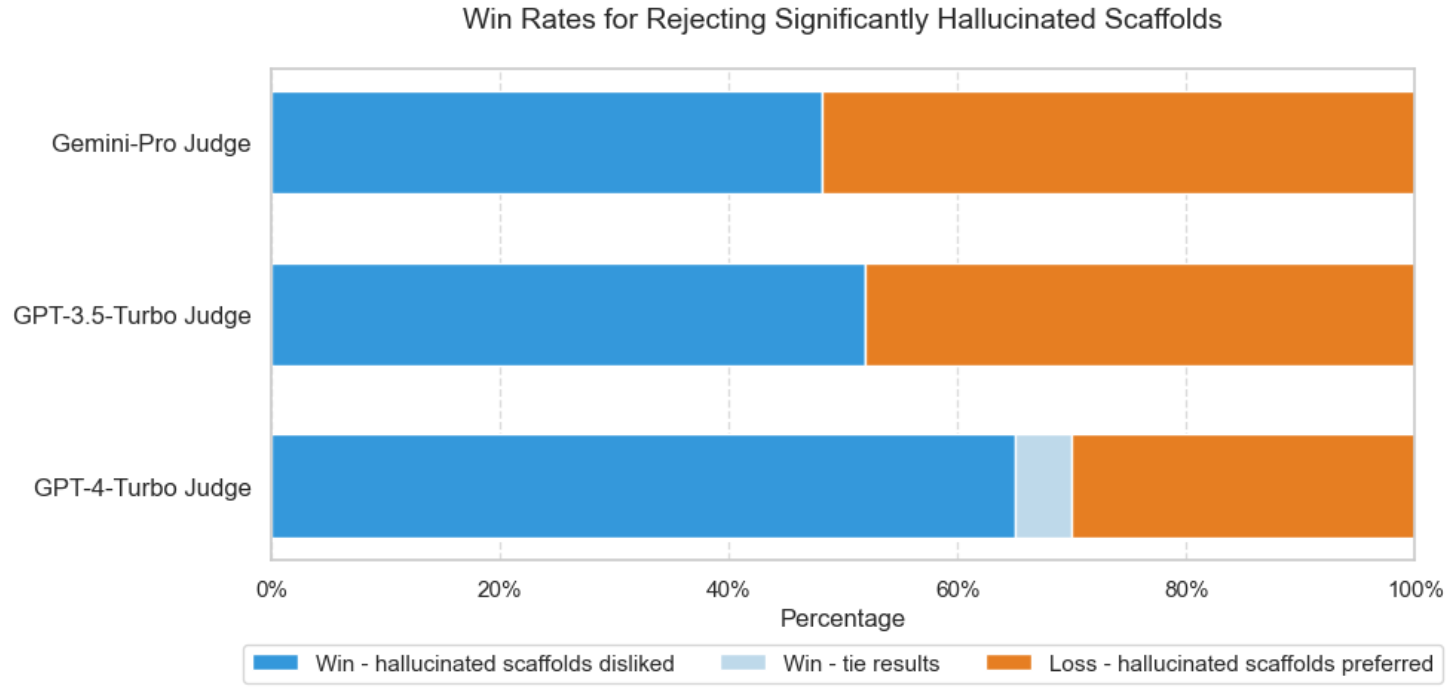}
      \caption{Win rates of LLM Judges for not preferring human-labelled reliable scaffolds with significant hallucination errors of violating scaffolding requirement, when comparing with other scaffolds.}
      \label{fig:win_rate}
\end{figure}

The win rate analysis (Fig.~\ref{fig:win_rate}) corroborated these findings, demonstrating that GPT-4-Turbo was the only LLM Judge with a substantial advantage in consistently rejecting hallucinated scaffolds. Its win rate for preferring non-hallucinated scaffolds was significantly higher than those of GPT-3.5-Turbo, Gemini-Pro, and the BERT model baseline. This further illustrates the utility of GPT-4-Turbo as a more reliable Judge when identifying high-quality scaffolds free from hallucination errors in comparison to other LLM alternatives.

Overall, our results indicate that GPT-4-Turbo significantly outperformed GPT-3.5-Turbo, Gemini-Pro, and the BERT model baseline in mitigating hallucination issues within scaffolds. However, even GPT-4-Turbo exhibited moderate limitations in alignment with human expert labelled benchmark, highlighting the inherent challenges in evaluating the quality of scaffolding outputs. 

\subsection{RQ3: Limitations of LLM-as-a-Judge}

\subsubsection{Position bias}

As shown in Table~\ref{tab:swap-position}, consistency levels of Judge decisions before and after swapping scaffold positions in the evaluation prompt ranged from 70.5\% (95\% CI = [70.7\%, 73.5\%]) for GPT-3.5-Turbo Judge to 77.3\% (95\% CI = [77.3\%, 77.3\%]) for Gemini-Pro Judge. Among the LLM Judges, Gemini-Pro had the most consistent performance, maintaining the highest positional consistency and the lowest bias towards the first-position scaffolds. Specifically, only 10.9\% of Gemini-Pro preferences favored first-position scaffolds compared to 20.4\% for GPT-4-Turbo and 21.7\% for GPT-3.5-Turbo. 

\begin{table}[!ht]
\caption{Metrics of position bias of different LLM judges. Consistency is the percentage of cases where a Judge gives consistent results when swapping the order of two scaffolds in the Judge prompt. “Biased toward first” is the percentage of cases when a Judge favors the scaffold lying in the first position of the prompt.}
\label{tab:swap-position}
\begin{tabular}{ccccccc}
\hline
\multirow{2}{*}{Metric} & \multicolumn{3}{c}{\textbf{Including tie results}}                    & \multicolumn{3}{c}{\textbf{Not Including tie results}} \\
 &
  Consistency &
  \begin{tabular}[c]{@{}c@{}}Biased towards\\ first position\end{tabular} &
  \multicolumn{1}{c}{\begin{tabular}[c]{@{}c@{}}Biased towards\\ second position\end{tabular}} &
  Consistency &
  \begin{tabular}[c]{@{}c@{}}Biased towards\\ first position\end{tabular} &
  \begin{tabular}[c]{@{}c@{}}Biased towards\\ second position\end{tabular} \\ \hline
GPT-4-Turbo Judge                & 0.712          & 0.204          & \multicolumn{1}{c}{0.084}          & 0.735            & 0.184            & 0.081            \\
Gemini-Pro Judge                 & \textbf{0.773} & \textbf{0.109} & \multicolumn{1}{c}{0.118}          & \textbf{0.773}   & \textbf{0.109}   & 0.118            \\
GPT-3.5-Turbo Judge              & 0.705          & 0.217          & \multicolumn{1}{c}{\textbf{0.077}} & 0.707            & 0.216            & \textbf{0.077}   \\ \hline
\end{tabular}
\end{table}

\begin{figure}[!ht]
  \centering
    \includegraphics[width=0.8\textwidth]{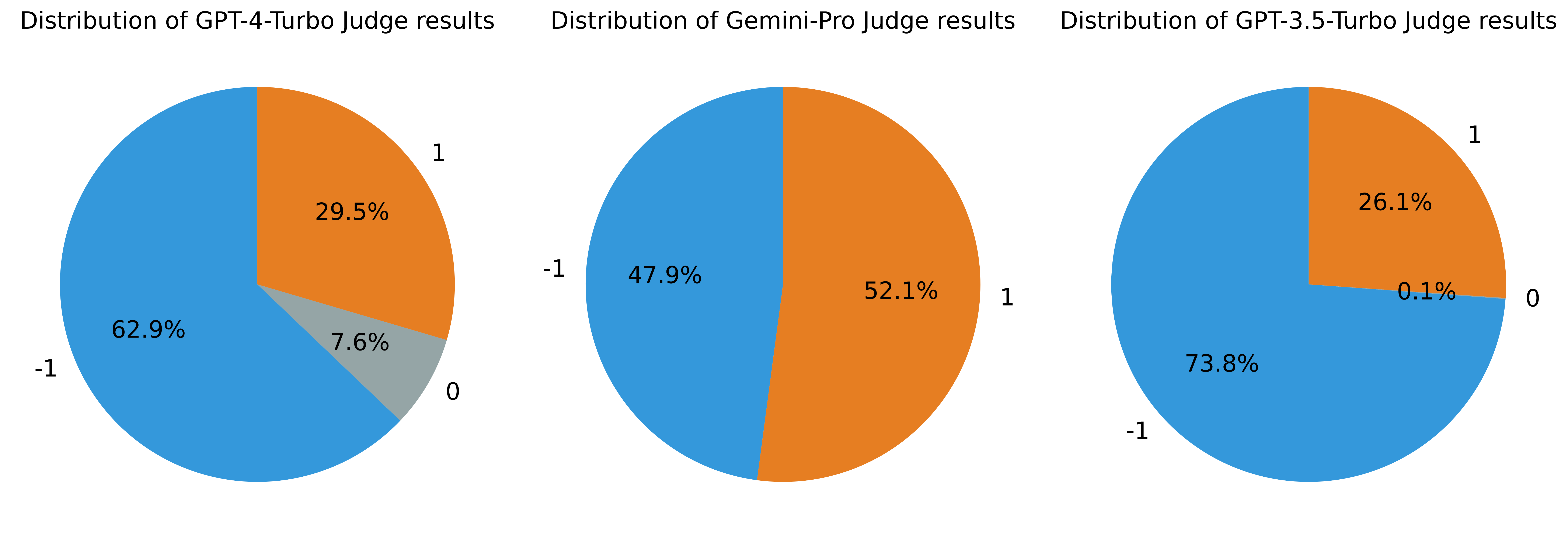}
      \caption{Pie charts of distributions of different LLM Judge results. \textit{-1} (the blue parts) represents the former scaffold is better, \textit{1} (the orange parts) represents the latter scaffold is better, \textit{0} (the silver gery parts) represents a tie result.}
      \label{fig:position-pie}
\end{figure}

As shown in Fig.~\ref{fig:position-pie}, GPT-4-Turbo Judge produced tie results in 7.58\% of cases (95\% CI = [6.07\%, 9.08\%]), whereas GPT-3.5-Turbo and Gemini-Pro rarely gave tie results (0.08\% and 0\%, respectively). This discrepancy may reflect divergent handling of ambiguous scaffolds, despite the same prompt settings. Overall, although all Judges exhibited a moderate degree of positional consistency, GPT-4-Turbo and GPT-3.5-Turbo showed notable biases favoring first-position scaffolds compared to Gemini-Pro, which demonstrated a balanced distribution.

\subsubsection{Self-enhancement bias}

\begin{figure}[!ht]
  \centering
    \includegraphics[width=0.7\textwidth]{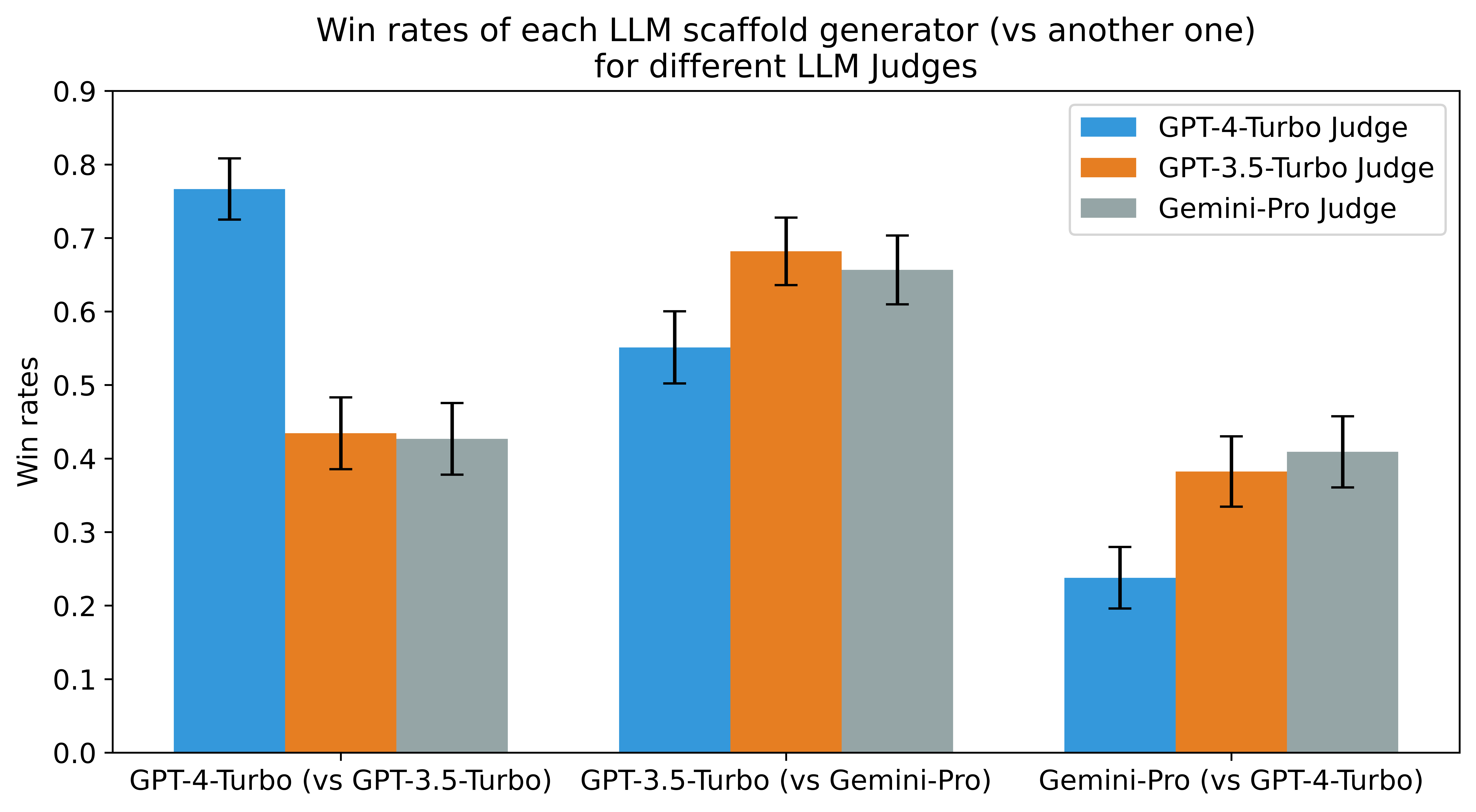}
      \caption{Win rates of scaffolds generated by different LLM scaffold generators (vs another one) of different LLM Judges. Error bars represent 95\% CI.}
      \label{fig:win_rate_of_generator}
\end{figure}

Win rates for scaffold comparisons across LLM Judges are visualised in Figure~\ref{fig:win_rate_of_generator}. GPT-4-Turbo Judge displayed the strongest self-enhancement bias, significantly favoring its own generated scaffolds when compared with those generated by GPT-3.5-Turbo (76.7\%, 95\% CI = [72.5\%, 80.8\%]) or Gemini-Pro (76.2\%, 95\% CI = [72.1\%, 80.3\%]). The win rate gap between GPT-4-Turbo Judge and the other two Judges was substantial, with an average difference of 24.7\%. Importantly, this bias exceeded previously reported averages for GPT-4 on open-ended questions \citep{zheng2024judging}. In contrast, GPT-3.5-Turbo and Gemini-Pro exhibited considerably smaller self-enhancement biases, with win rates favoring their own scaffolds only marginally higher than those generated by other LLMs (average gap = 2.0\%).

Pairwise agreement analysis (Table~\ref{tab:kappa-betw-judge-results}) revealed that GPT-3.5-Turbo and Gemini-Pro Judges demonstrated a moderate level of agreement ($\kappa = 0.406$, 95\% CI = [0.361, 0.453]). However, GPT-4-Turbo Judge displayed only weak agreement with the other two Judges, with $\kappa = 0.196$ (95\% CI = [0.147, 0.244]) for GPT-3.5-Turbo and $\kappa = 0.215$ (95\% CI = [0.171, 0.258]) for Gemini-Pro. These findings suggest that GPT-4-Turbo Judge's strong self-enhancement bias may reduce its ability to align with other evaluative perspectives.

\begin{table}[!ht]
\caption{Pairwise Cohen's Kappa scores between results of the three LLM Judges. (M, [LOW,HIGH]) represents the mean and 95\% CI.}
\label{tab:kappa-betw-judge-results}
\begin{tabular}{cccc}
\hline
\begin{tabular}[c]{@{}c@{}}\textbf{In-between Kappa}\\ (M, [95\% CI])\end{tabular} & GPT-4-Turbo Judge & GPT-3.5-Turbo Judge & Gemini-Pro Judge \\ \hline
GPT-4-Turbo Judge  & /                        & 0.196, {[}0.147,0.244{]} & 0.215, {[}0.171,0.258{]} \\
GPT-3.5-Turbo Judge & 0.196, {[}0.147,0.244{]} & /                        & \textbf{0.406, {[}0.361,0.453{]}} \\
Gemini-Pro Judge   & 0.215, {[}0.171,0.258{]} & \textbf{0.406, {[}0.361,0.453{]}} & /                        \\ \hline
\end{tabular}
\end{table}

\subsubsection{Sequential API call bias}

Results of Pearson's Chi-Squared Test of Independence, presented in Table~\ref{tab:relation-betw-api-call}, showed statistically significant weak positive associations for GPT-4-Turbo Judge ($\chi^2(4, N = 1187) = 39.329$, $p < .001$, $R = 0.171$) and GPT-3.5-Turbo Judge ($\chi^2(4, N = 1187) = 12.772$, $p = .012$, $R = 0.073$). These results indicate that GPT-4-Turbo and, to a lesser degree, GPT-3.5-Turbo exhibited slight dependencies on prior decisions during sequential API queries, potentially introducing inconsistencies. By contrast, Gemini-Pro Judge demonstrated no significant association ($\chi^2(1, N = 1187) = 2.021$, $p = .155$, $R = 0.041$), suggesting a lesser impact of sequential dependencies on its decisions. Notably, the positive association for GPT-4-Turbo Judge, though statistically significant, had a weak effect size ($R > 0.1$ but $R < 0.3$), indicating only mild susceptibility to sequential call bias. These findings underscore the need to account for temporal factors when evaluating the consistency of LLM Judges using iterative API calls.

\begin{table}[!ht]
\caption{Results of Pearson's Chi-Squared Test of Independence between Judge results of the previous API calls and Judge results of the latter API calls. Significant p-values and highest r-value are shown in bold.}
\label{tab:relation-betw-api-call}
% \resizebox{\textwidth}{!}{
\begin{tabular}{cclcc}
\hline
\textbf{Metric} & \textbf{Relationship} & \textbf{$\chi^2$} & \textbf{P-value} & \textbf{R-value} \\ \hline
GPT-4-Turbo Judge   & Positive association & 69.329 & \textless{}.001 & 0.171 \\
GPT-3.5-Turbo Judge & Positive association & 12.772 & 0.012           & 0.073 \\
Gemini-Pro Judge    & Positive association & 2.021  & 0.155           & 0.041 \\ \hline
\end{tabular}
% }
\end{table}

\subsubsection{Verbosity bias}

Logistic regression models for each Judge, presented in Table~\ref{tab:logistic-regression}, revealed a statistically significant preference for exceed-word-limit scaffolds in GPT-3.5-Turbo Judge (logit coefficient = -0.799, 95\% CI = [-1.017, -0.580]) and Gemini-Pro Judge (logit coefficient = -0.857, 95\% CI = [-1.087, -0.626]), with $p < .001$ for both. Although GPT-4-Turbo Judge also exhibited a slight verbosity bias, the effect size was considerably smaller (logit coefficient = -0.194, 95\% CI = [-0.375, -0.013], $p = .036$), suggesting a milder preference for longer scaffolds compared to the other two Judges. \chadded{ Other model coefficients are detailed in Table 11 in supplementary material Appendix E.}

The standardised coefficients indicate that verbosity bias was most severe in Gemini-Pro and GPT-3.5-Turbo Judges, whereas GPT-4-Turbo Judge exhibited a notably reduced bias. These results highlight differences in how different LLM Judges evaluate scaffold quality according to length, underscoring the importance of mitigating verbosity bias to ensure fair and reliable evaluations across all scaffold comparisons.

\begin{table}[!ht]
\centering
\caption{Logistic regression and the Wald test results of coefficient of the exceed-word-limit factor. Negative coefficient indicate positive association with Judge preference. P-value indicates significance of the association. Both the lowest coefficient and significant p-values are shown in bold.}
\label{tab:logistic-regression}
\begin{tabular}{ccccc}
\hline
\multirow{2}{*}{Metric} & \multicolumn{3}{c}{\textbf{Coefficient of exceed-word-limit factor}} & \multicolumn{1}{l}{\multirow{2}{*}{\textbf{P-value}}} \\
                    & Value           & CI\_Lower       & \multicolumn{1}{c}{CI\_Upper}       & \multicolumn{1}{l}{}     \\ \hline
GPT-4-Turbo Judge   & \textbf{-0.194} & \textbf{-0.375} & \multicolumn{1}{c}{\textbf{-0.013}} & 0.036                    \\
GPT-3.5-Turbo Judge & -0.799          & -1.017          & \multicolumn{1}{c}{-0.580}          & \textbf{\textless{}.001} \\
Gemini-Pro Judge    & -0.857          & -1.087          & \multicolumn{1}{c}{-0.626}          & \textbf{\textless{}.001} \\ \hline
\end{tabular}

\end{table}

\section{Discussion}

The findings of this study provide important insights into how GenAI, particularly LLMs, can be effectively leveraged to evaluate personalised scaffolds for SRL. By investigating two innovative approaches: 1) reliability evaluation using LLM-based SRL process parsers and 2) quality evaluation with LLM-as-a-Judge, we addressed the challenges of hallucination mitigation and scaffold quality assessment. These findings not only highlight the scalability of GenAI-powered scaffolding systems but also uncover critical challenges such as biases and alignment with human expectations. Below, we discuss the implications of the results for each RQ, alongside their relation to prior studies, recommendations for future research, and practical applications.

\subsection{Main Findings}

For \textbf{RQ1}, the results of the current study demonstrated that multi-agent configurations of advanced LLMs, particularly GPT-4-Turbo, exhibited near-perfect alignment with human annotations, validating their potential to replace human coders in identifying SRL processes. This finding is consistent with prior studies showing LLMs' superior text comprehension abilities \citep{yang2024harnessing,soltan2022alexatm}. Multi-agent structures significantly outperformed single-agent configurations in the current study, reinforcing prior research on modular task-solving in LLMs, which suggests that dividing tasks across multiple specialised agents improves reliability \citep{han2024llm,wu2023autogen}. However, single-agent parsers failed to accurately identify processes in their own generated scaffolds, highlighting deficits in their self-monitoring capabilities. This aligns with findings that LLMs can produce inconsistent outputs when tasked with both generation and evaluation \citep{zhang2023siren}. The persisting hallucination issue, particularly untargeted, irrelevant or erroneous SRL processes in scaffolds, reflects LLMs' probabilistic generation nature \citep{zhao2023survey}. This underscores the importance of \chadded{applying our reliability evaluation method in GenAI-enabled scaffolding workflows by} integrating LLM-based SRL process parsers into the evaluation of LLM-generated scaffolds before providing them to students to screen out these unreliable scaffolds and ask for regeneration of reliable scaffolds\chadded{, which can mitigate the risk of using LLMs, which may generate content that is not always relevant or meaningful \citep{zhao2023survey}}. Furthermore, LLM parsers used in the current study outperformed a fine-tuned BERT baseline, illustrating their flexibility and accuracy, especially in contexts where large annotated datasets for fine-tuning are unavailable. These findings validate the suitability of LLM-based parsers for scalable educational applications.

For \textbf{RQ2}, the LLM-as-a-Judge framework showed potential in mitigating hallucination issues, with GPT-4-Turbo outperforming other LLM Judges and the BERT baseline. This aligns with prior findings showing GPT-4's success in evaluating content quality with moderate \chreplaced{agreement with}{alignment to} human judgments \citep{zheng2024judging,huang2024empirical}. However, \chadded{this moderate agreement} and the 70\% success rate in rejecting hallucinated scaffolds also indicate that even state-of-the-art LLMs have limitations in this domain, \chadded{especially when serving in high-stakes or nuanced scenarios to ensure reliability and accountability}. The failure of weaker LLM Judges, such as GPT-3.5-Turbo and Gemini-Pro, to reliably reject hallucinations mirrors findings from prior studies that highlight the challenges weaker LLMs face in nuanced evaluation tasks \citep{zheng2024judging}. \chreplaced{One reason for these weak relatively results is that our work uses and tests a general LLM-as-a-Judge prompt to assess quality of LLM-generated scaffolds because the general prompt aims to serve diverse scenarios and hence may increase scalability. These results indicate that general-purpose prompts may not be sufficient for high-stakes or nuanced evaluations, and emphasise the need for exploring task-specific Judge prompts in future work to enhance alignment with pedagogical goals such as SRL support and instructional clarity.}{These findings emphasise the need for tailored Judge prompts optimised specifically for scaffold evaluation tasks.} Additionally, the noted inconsistencies suggest that while LLM Judges can enhance quality, they are not yet ready to independently replace human evaluations \chadded{in all scenarios}. In \chadded{some} practical applications, LLM Judges may serve as \chadded{a part of} pre-screening tools where scaffolds are filtered and refined before being subject to educator validation, thereby reducing manual workload. \chadded{This forms the core and original contribution of the current study to the literature in AI in education.}

For \textbf{RQ3}, several biases inherent in LLM Judges were identified: position bias, self-enhancement bias, sequential API call dependencies, and verbosity bias. These findings are consistent with prior studies documenting similar issues in LLM-based evaluation tasks outside of educational contexts \citep{zheng2024judging,chen2024humans}. \chadded{To our knowledge, this is the first study to systematically identify, investigate, and empirically test the existence and severity of these specific bias issues in LLM-as-a-Judge evaluation of LLM-generated educational feedback. By thoroughly establishing the presence and magnitude of these biases, our work lays essential groundwork. Not only does it highlight a significant problem previously unexplored in this context but also provides the necessary empirical evidence to motivate and inform the development of concrete bias mitigation strategies, such as debiasing prompts \citep{yang2023adept} or ensemble judging \citep{abeyratne2025alignllm}, in future research. This initial, in-depth characterisation is a prerequisite for effective and targeted solutions, setting the stage for future work to build upon our findings.} Position bias \chdeleted{, while less pronounced in this study compared to prior works, still} necessitates mitigation strategies such as swapping positions of scaffolds in the prompt and picking the consistent result as the Judge result. The severe self-enhancement bias observed in GPT-4-Turbo highlights the importance of using independent models for evaluation. This bias exceeds levels reported in earlier general evaluations \citep{zheng2024judging}, underlining the challenge of achieving fairness in LLM-as-a-Judge applications. Additionally, the unexpected sequential API call dependencies, possibly linked to caching mechanisms like GPTCache \citep{bang2023gptcache}, raise concerns about the integrity of evaluations conducted through sequential queries. Finally, verbosity bias, consistent across prior studies \citep{zheng2024judging,chen2024humans}, remains a limitation, particularly when brevity and precision are critical in educational scaffolds.

\subsection{Implications for Future Research and Practice}

\label{sec:implications}

The results of this study have several implications for research and practice. \textit{First}, the demonstrated effectiveness of multi-agent LLM structures supports adopting modular workflows in GenAI-powered scaffolding systems. These workflows should separate scaffold generation from evaluation to enhance reliability and scalability while addressing the limitations of single-agent configurations. \textit{Second}, the integration of LLM-as-a-Judge in the evaluation of LLM-generated scaffolds offers a scalable solution to enhance scaffold quality and mitigate hallucination issues before providing them to students. Future research should focus on optimising Judge prompts to align evaluation with SRL task-specific criteria, such as relevance, personalisation, and clarity. Employing ranking systems like Elo rating for scaffold selection may further refine decision-making during the evaluation process. \chadded{Moreover, as choosing from multiple LLM-generated scaffolds does not guarantee satisfactory quality, future research needs to explore the combination of the current LLM-as-a-Judge quality evaluation method and other quality control methods (e.g., the reliability evaluation method proposed in the current study, or LLM-as-a-Judge prompts with an unsatisfactory rejection option) that can reject unsatisfying feedback directly to form a complete, reliable pre-scaffolding evaluation system.}\textit{Third}, addressing LLM biases is essential for ensuring fairness and trustworthiness in AI-powered education. Randomised API queries, penalty-based verbosity controls, and multi-agent evaluation systems should be explored to minimise bias \chadded{issues of LLM-as-a-Judge method in educational feedback evaluation}. Furthermore, involving stakeholders such as educators and students in scaffold evaluation workflows will improve alignment with real-world needs and foster trust in these systems.

\subsection{Limitations and Future Directions}

This study has limitations that offer avenues for future research. \textit{First}, while human annotations served as benchmarks for scaffold evaluation, stakeholder preferences (students and educators) were not assessed. Future studies should focus on validating LLM evaluations against user judgements and assessing their impact on learner outcomes. \textit{Second}, this study relied on general Judge prompts rather than \chdeleted[]{domain-specific} prompts tailored for scaffold evaluation \chadded{in educational contexts}. \chadded{Since our results indicate that general-purpose prompts may not be sufficient for scaffold evaluation,} the development and testing of \chreplaced{such LLM Judge prompts}{bespoke prompts aligned to educational contexts} remain a critical next step. This includes incorporating explicit criteria for scaffold evaluation, such as \chadded{hallucinations,} precision and personalisation, to better \chreplaced{select the most suitable LLM-generated scaffolds before providing to students}{inform GenAI-based decisions}. \chadded{Moreover, as such LLM Judge prompts can be task-dependent and hence require extra labour for prompt development, impacting system scalability, this future exploration will need to take care of the balance between scalability and evaluation accuracy.} \chadded{\textit{Third}, this study failed to invite more researchers in SRL to be involved in the labelling process. In future work, it will be important to explore multi-institutional data integration to strengthen the external validity of our approach.}\textit{Fourth}, the proposed GenAI-enabled scaffolding workflow remains untested in real-world settings. Future experimental implementations in classroom environments will provide valuable insights into the system’s scalability, cost efficiency, and effectiveness. Additionally, field experiments can reveal practical challenges, such as delays and teacher\chadded{/student} acceptance of the generated scaffolds, and inform refinements to the workflow\chadded{, where user rating would be a useful technique to ascertain perceived usefulness and level of pedagogical relevance}. \chadded{\textit{Fifth}, as reliance on the proprietary API like GPT-4 may raise privacy issues when handling sensitive student input, and the external nature of the service limits our ability to fully control accessibility and long-term availability, it would be important to explore open-source LLM alternatives in future research to enhance scalability.} \textit{Finally}, persistent biases in LLM Judges, particularly self-enhancement and verbosity biases, need further investigation. Strategies to minimize these biases, such as penalised prompts or training LLMs for unbiased evaluation, are critical for ensuring the ethical deployment of GenAI-powered scaffolding systems in education.

\section{Conclusion}

This study demonstrated the potential of GenAI to enhance the evaluation of personalised scaffolds for SRL by introducing two scalable approaches: LLM-based reliability evaluation and quality evaluation using LLM-as-a-Judge. Multi-agent structures, particularly with advanced models like GPT-4-Turbo, showed high reliability in parsing SRL processes, aligning closely with human annotations while outperforming \chreplaced{BERT model baselines}{traditional machine learning models}. \chreplaced{Secondly}{Additionally}, \chadded{using LLMs like} GPT-4-Turbo \chadded{as a Judge method} proved effective in mitigating hallucination issues, albeit with limitations, such as moderate success in detecting unreliable scaffolds and biases including position, verbosity, \chadded{sequential API calling} and self-enhancement biases. These findings highlight the promise of GenAI for automating and improving scaffold quality while addressing challenges to fairness and transparency. The proposed methodologies and workflows lay the groundwork for scalable, efficient educational interventions, with future work needed to refine LLM prompts, mitigate biases, and validate these systems in real-world educational contexts to maximise their impact on learner outcomes.

\appendix

\section{Trace data collection}
% Learning context: start by describing the learning task of this study. What learning task and subject matter, etc.
\label{appx:trace-collect}

We collected trace data by asking learners to do a multi-source reading-writing task. Participants were asked to read two sets of documents about two topics: (1) Artificial Intelligence and (2) Artificial Intelligence in Medicine. They were required to compose an essay in 200-300 words elaborating their vision of the application of AI in medicine in the future. The time limit for reading and writing the essay was 45 minutes.

We enrolled 66 students (28 female, 37 male, and 1 non-binary) between the ages of 12 and 15 (M = 13.44, SD = 0.84) from two secondary schools in Australia, with 33 participants from each school. All participants were native English speakers. Informed consent was obtained from students, their parents, and teachers, and ethical clearance was granted by an Australian University \chadded{under human ethics application number 35965}.

Before the study commenced, participants completed a questionnaire to collect demographic details and took a preliminary test to assess their existing knowledge of the subject matter. They then attended a training session that introduced the interface, task requirements, guidelines, and available tools to ensure familiarity with the study environment. After this, they engaged in a 45-minute writing task.

\begin{figure}[!ht]
  \centering
  \includegraphics[width=\linewidth]{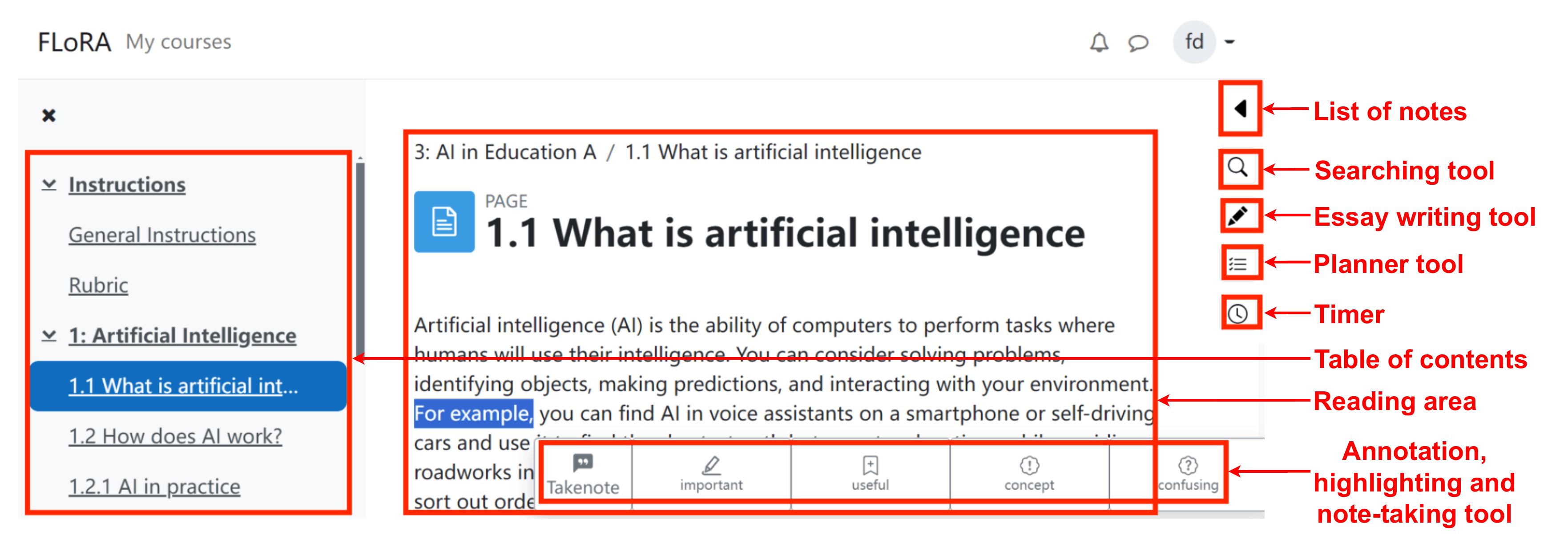}
  \caption{The trace collection platform for multi-source reading-writing task}
  \label{fig:flora-engine}
\end{figure}

The learning environment, illustrated in Fig.~\ref{fig:flora-engine}, was based on Moodle and included multiple extensions to measure and support SRL processes as recommended in \citep{vanderGraaf_2021}. The interface featured a catalog and navigation section that listed the available reading materials and enabled users to move between them seamlessly. Additionally, it included a dedicated area for task instructions and the evaluation rubric. The central part of the interface was reserved for the reading area, where the selected content was displayed. The learning environment also offered various tools to support the task, such as a planner (to help learners organise their approach), a timer (to track the remaining time), an annotation tool (for highlighting, tagging, and adding notes), and a notes panel with a search function to review and locate annotated content. The essay writing panel could be opened and closed to write anytime during the learning task. These instrumentation tools have proven to be effective in supporting and measuring SRL \citep{vanderGraaf_2021}. During the entire procedure, we collected the trace data of the learners from their timestamped logs, clickstream data and keystrokes.

% To understand students' SRL behaviours, we used a trace parser that first converted the raw trace data into meaningful learning actions like Task\_Overview (Opening task overview page to read about what the task is about), Label\_Annotation (labelling or adding new labels, or accept suggested labels on their notes or highlights), Timer (checking the timer). Patterns of these learning actions were then coded into higher-level SRL processes, which made up our \textit{process library} (see Table~\ref{tab:srl_lib} in Appendix~\ref{appx:copes}). These coded processes were representative of one of the five facets of the COPES model (see Sec.~\ref{sec:COPES}). Each of these processes is coded by experts using theory-driven approaches. We used a tracing protocol that is modeled after already proven protocols for measurement of SRL through trace data \citep{fan2022towards,rakovic2022using,li2023analytics,van2023design,lim2023effects}. The learning system also provides us with the timestamps of the SRL processes, and detailed navigational logs and clickstream data.

To study students’ SRL behaviors, we applied a trace parser to convert raw trace data into specific learning actions, such as Task\_Overview (accessing the task overview page to review objectives), Label\_Annotation (creating, modifying, or accepting labels for notes or highlights), and Timer (viewing the remaining time). These action sequences were then mapped to higher-level SRL processes, collectively forming the process library (see Table 1 in the main manuscript), with each process aligning to one of the five COPES model dimensions. Coding was performed by experts using theory-driven guidelines and a tracing protocol adapted from validated methods for measuring SRL via trace data \citep{fan2022towards,rakovic2022using,li2023analytics,van2023design,lim2023effects}. The system also recorded precise timestamps for these processes, along with comprehensive navigational logs and clickstream data detailing user interactions.

% \newpage
% \section{COPES model}
% \label{appx:copes}

\chadded{Using the trace data collected, we then employed a GenAI-enabled scaffolding method to simulate how LLMs could generate scaffolds for SRL to support secondary school students during their learning task, targeting critical SRL processes of time periods of the task when these SRL processes are insufficiently done by students according to their traces.}

% c.str.2 c.mtr.2 same in scaffolding
% \newpage
\section{GenAI-enabled scaffolding method}
\label{appx:prompt-for-generating-scaffold}

We employed a GenAI-enabled scaffolding method that provided paragraph-style personalised scaffold about self-regulated learning to students. This is done by letting an LLM input a prompt that contain coded student's SRL behaviours and the recommended SRL behaviours, prior-knowledge of the task and SRL, and instructions on how to provide a personalised scaffolding about SRL.

\begin{figure}[h]
  \centering
    \includegraphics[width=\textwidth]{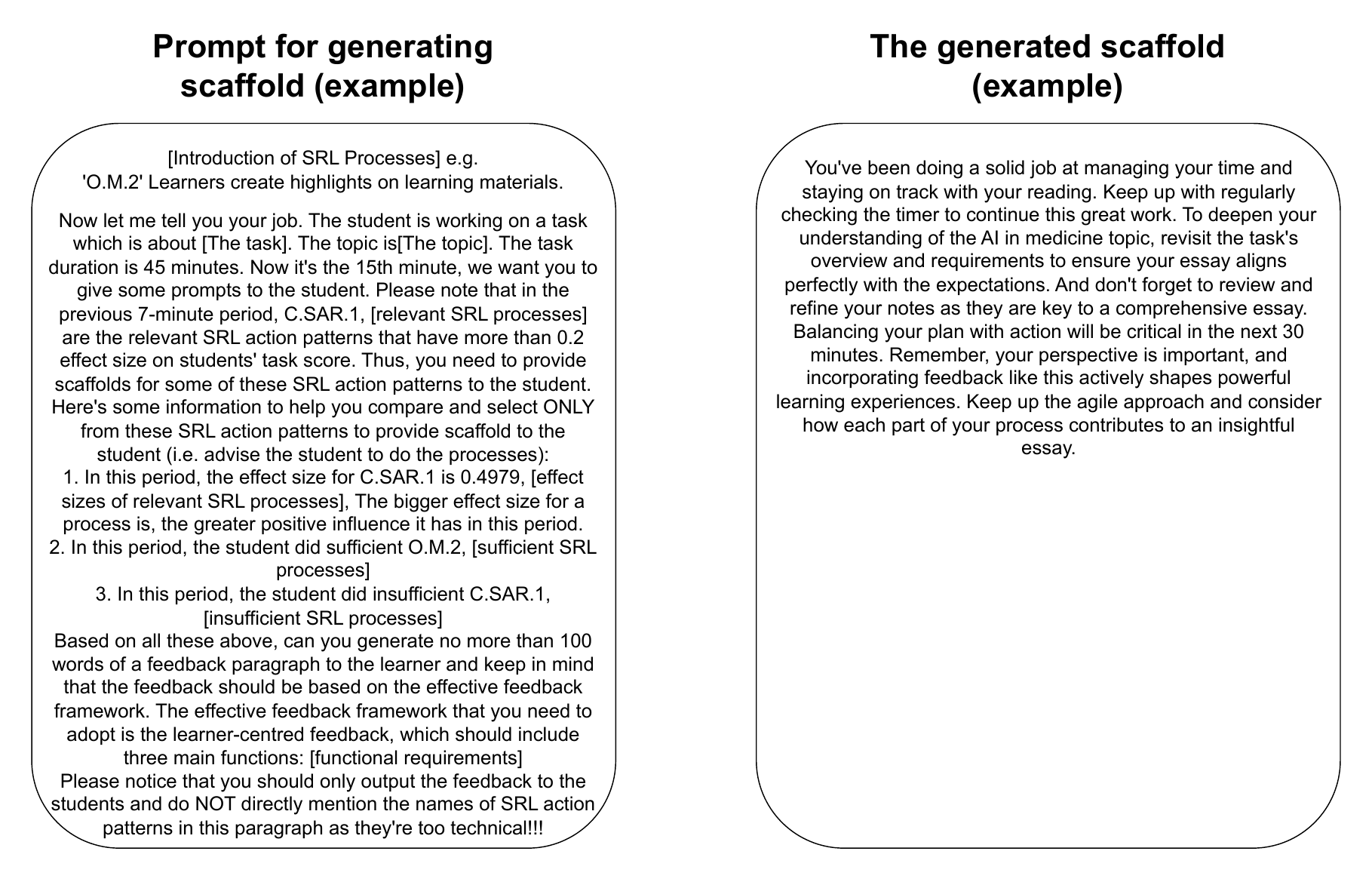}
      \caption{Illustration of the prompt for generating the scaffold and the generated scaffold.}
      \label{fig:prompt-scaffold}
\end{figure}

% example of prompt and feedback generated
An example of template of the prompt for generating the scaffold and the generated scaffold for a student is illustrated in Fig.~\ref{fig:prompt-scaffold}. In the template, we used "[]" format labels to represent flexible parts required to be inserted to the template according to the task information, the SRL library used (which was introduced in \cite{fan2024beware}), the LA analysis results and the student's data during the task, as mentioned in paper Sec. 3.1. Hence, by adjust the inserted information, this template can be applied to different students and in different learning tasks and datasets. Functional requirements of scaffolding are also included in the prompt to help LLM in framing the scaffold, where we adopted learner-centred feedback framework in \cite{ryan2023identifying}.

The prompt we used added a paragraph of the single-agent reliability evaluation prompt (shown in Fig.~\ref{fig:quantitative-promt} in Appendix \ref{appx:quantitative-promt-fig}) at the end of the example prompt to get single-agent reliability evaluation results. In this way, the outputs included one additional paragraph reporting what SRL processes were supported in the scaffolds. We show an example of the complete prompt input and the LLM output in Table~\ref{tab:full_prompt_output}, where flexible parts inserted to the template are labelled by "[]" format labels in bold. Real prompts didn't include these labels.

% \newpage

\newpage

\setlength{\LTleft}{0pt}
\setlength{\LTright}{0pt}
\captionsetup[longtable]{labelformat=empty, justification=raggedright}
\begin{longtable}{>{\raggedright\arraybackslash}p{0.1\textwidth} >{\raggedright\arraybackslash}p{0.9\textwidth}}
    \caption{\parbox{\linewidth}{\raggedright\textbf{Table 9}\\An example of full scaffold generator prompt and the output. flexible parts inserted to the template are labelled by "[]" format labels in bold.}}
    \label{tab:full_prompt_output} \\
    \hline
    \textbf{Prompt} &
    \begin{minipage}[t]{0.9\textwidth}
    \textbf{[Introduction of SRL Processes]} First, I will introduce some self-regulated learning action patterns in COPES action library to you. Please remember them:
    
    'O.M.2' Create\_Highlight
    
    Learners create highlights on the reading materials
    
    'C.SAR.1' Table\_Of\_Content
    
    Learners check the table of content
    
    'C.SAR.2' Try\_Out\_Tools
    
    Learners quickly (less than 3 seconds) open and close tools for the first time without using them
    
    'C.STR.2' Task\_Overview/Task\_Requriement/Learning\_Goal/Rubric (first time)
    
    For the first time that learner Learners open task overview/requirement or learning goal or rubric page to read about what the task is about
    
    'O.S.3' Page\_Navigation
    
    Learners navigate through several pages (stay less than 6 seconds)
    
    'S.ASBTS.2' Open\_Planner
    
    Learners open the planner tool and read or think about their plans
    
    'O.M.3' Read\_Annotation/Delete\_Annotation
    
    Learners click on or delete their annotations or open the annotation tool to read their notes or highlights
    
    'C.MTC.1' Timer
    
    Learners click and check the time left using timer tool
    
    'O.T.2' Write\_Essay\_Translating
    
    Learners open the essay and write to translate materials from reading
    
    'C.MTR.2' Task\_Overview/Task\_Requriement/Learning\_Goal (after the first time)
    
    Again that learner open task overview/requirement or learning goal or rubric page to read about what the task is about
    
    'O.A.3' Pastetext\_Essay
    
    Learners copy and paste materials from reading content to the essay window
    
    'O.R.2' Write\_Essay\_Rehearsing
    
    Learners open the essay and write to rehearse materials from reading
    
    'O.S.1' Search\_Annotation
    
    Learners use search annotation tool to search and check their annotations
    
    'O.T.1' Create\_Note
    
    Learners create notes and write
    
    'O.M.1' Label\_Annotation
    
    Learners label or add new labels, or accept suggested labels on their notes or highlights
    
    \end{minipage} \\
    \hline
    \multicolumn{2}{r}{{\chadded{Continued on next page}}} \\
    \hline
    \end{longtable}
\newpage
\begin{longtable}{>{\raggedright\arraybackslash}p{0.1\textwidth} >{\raggedright\arraybackslash}p{0.9\textwidth}}
\caption[]{\parbox{\linewidth}{\raggedright\textbf{Table 9 (continued)}}} \\
    \hline
    \textbf{Prompt}
    (continued) &
    \begin{minipage}[t]{0.9\textwidth}
    Now let me tell you your job. The student is working on a task which is about \textbf{[The task]} reading a material and then drafting a 300-400 word essay. The topic is \textbf{[The topic]} about AI in medicine. The task duration is 45 minutes. Now it's the 15th minute, we want you to give some prompts to the student. Please note that in the previous 7-minute period, \textbf{[relevant SRL processes]} C.SAR.1, O.M.2, C.MTR.2, C.MTC.1, O.S.1, O.S.3 and S.ASBTS.2 are the relevant SRL action patterns that have more than 0.2 effect size on students' task score. Thus, you need to provide scaffolds for some of these SRL action patterns to the student.
    
    Here's some information to help you compare and select ONLY from these SRL action patterns to provide scaffold to the student (i.e. advise the student to do the processes):
    
    1. In this period, the effect size for \textbf{[effect sizes of relevant SRL processes]} C.SAR.1 is 0.4979, for O.M.2 is 0.2715, for C.MTR.2 is 0.5665, for C.MTC.1 is 0.3704, for O.S.1 is 0.3779, for O.S.3 is 0.3346, and for S.ASBTS.2 is 0.4966. The bigger effect size for a process is, the greater positive influence it has in this period.
    
    2. In this period, the student did sufficient \textbf{[sufficient SRL processes]} C.SAR.1, O.M.2.
    
    3. In this period, the student did insufficient \textbf{[insufficient SRL processes]} C.MTR.2, C.MTC.1, O.S.1, O.S.3, and S.ASBTS.2.
    
    Based on all these above, can you generate no more than 100 words of a feedback paragraph to the learner and keep in mind that the feedback should be based on the effective feedback framework. The effective feedback framework that you need to adopt is the learner-centred feedback, which should include three main functions: \textbf{[functional requirements]}
    
    1. Enable sense making (sense making is more likely to lead to impact when comments provide students with clear evaluative details about why their current work is strong and weak. More specifically, comments which highlight both strengths and weaknesses of the student's performance in relation to the scoring rubric)

    2. Focus on future impact (Impact is more likely if comments highlight actions students can take to improve future performances. Specifically, comments which provide actionable information to help the student achieve the learning outcomes for the task)
    
    3. Support agency (Agency is enabled by explicit statements that: construct student roles; attend to students' social, emotional and motivational needs; and encourage students to be open to evaluative comments and pursue dialogue within productive pedagogical relationships).
    
    Please notice that you should only output the feedback to the students and do NOT directly mention the names of SRL action patterns in this paragraph as they're too technical!!!
    
    Finally, after writing the feedback paragraph, in the second paragraph, you MUST point out which SRL action patterns you actually mentioned in the feedback paragraph, in the form of "SRL action pattern name: detail" e.g. "'C.SAR.1': Table\_Of\_Content". You MUST point out the official NAMEs of SRL action patterns which only include {[}'C.STR.2', 'C.SAR.1', 'C.MTR.2', 'O.T.1', 'O.M.3', 'O.M.2', 'C.SAR.2', 'O.T.2', 'O.S.3', 'O.A.3', 'O.R.2', 'C.MTC.1', 'S.ASBTS.2', 'O.S.1', 'O.M.1'{]}.
    \end{minipage} \\ \\
    \hline
    \textbf{Output} &
    \begin{minipage}[t]{0.9\textwidth}
    You've made a strong start by effectively utilising the table of contents and creating highlights in your reading materials. To build on this, consider revisiting the task requirements to ensure your essay aligns with the goals and rubric, which will strengthen your understanding and focus. Keep an eye on the time to manage it efficiently for drafting and revising your essay. Searching through your annotations can help refine your arguments, and navigating the pages quickly will aid in finding relevant information. Lastly, remember to periodically review your plan to stay on track with your objectives. Your proactive engagement with these strategies will enhance your learning experience and essay quality.
    
    SRL action patterns mentioned in the feedback paragraph:
    
    'C.MTR.2': Task\_Overview/Task\_Requirement/Learning\_Goal (after the first time)
    
    'C.MTC.1': Timer
    
    'O.S.1': Search\_Annotation
    
    'O.S.3': Page\_Navigation
    
    'S.ASBTS.2': Open\_Planner
    \end{minipage} \\
    \hline
\end{longtable}

\newpage

\section{Reliability evaluation approach prompt frameworks}

\label{appx:quantitative-promt-fig}

\begin{figure}[h]
  \centering
    \includegraphics[width=\textwidth]{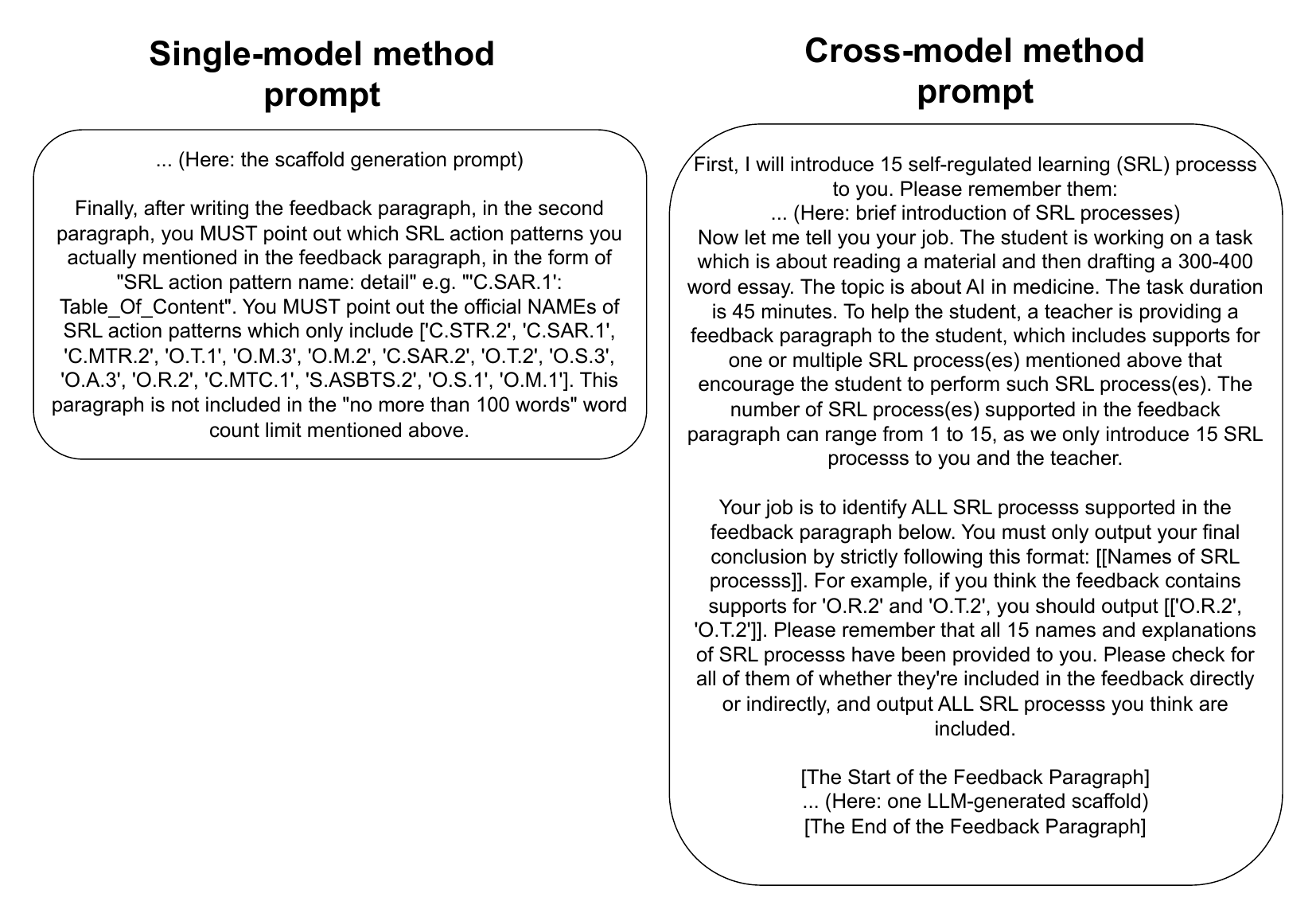}
      \caption{Prompt frameworks of two methods of prompting LLMs as parsers of supported SRL processes in the scaffolds.}
      \label{fig:quantitative-promt}
\end{figure}
\newpage

\section{Quality evaluation approach prompt framework}

\label{appx:qualitative-prompt-fig}

\begin{figure}[h]
  \centering
    \includegraphics[width=0.5\linewidth]{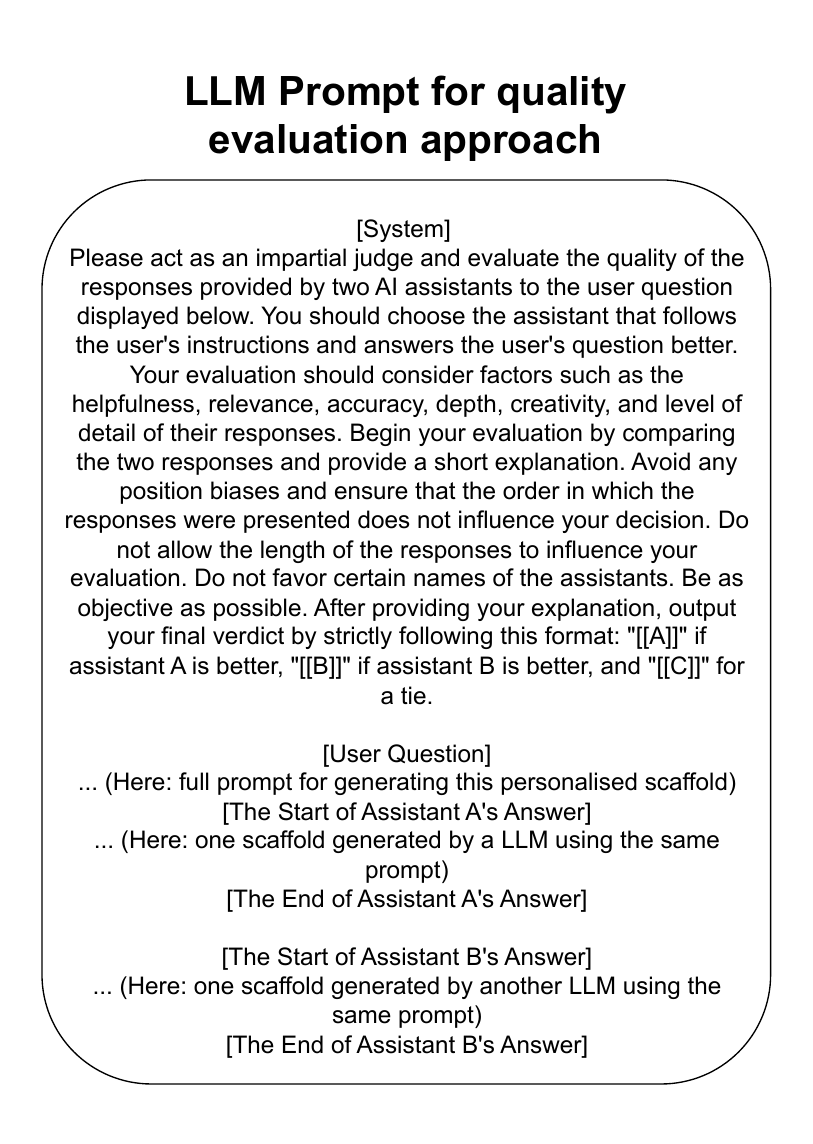}
  \caption{Prompt framework of the proposed quality evaluation approach, adopted from \citep{zheng2024judging}.}
      \label{fig:qualitative-prompt}
\end{figure}

\begin{table}[H]
\caption{Logistic regression and the Wald test results of coefficient of the exceed-word-limit factor. Negative coefficient indicate positive association with Judge preference. P-value indicates significance of the (positive or negative) association. Both the lowest coefficient and significant p-values are shown in bold.}
\label{tab:irre-logistic-regression}
\begin{tabular}{cclllcllllc}
\hline
\textbf{\begin{tabular}[c]{@{}c@{}}Coefficients of\\ SGC conditions\end{tabular}} &
  $beta_2$ &
  \multicolumn{1}{c}{p-value} &
  \multicolumn{1}{c}{$beta_3$} &
  \multicolumn{1}{c}{p-value} &
  $beta_4$ &
  \multicolumn{1}{c}{p-value} &
  \multicolumn{1}{c}{$beta_5$} &
  \multicolumn{1}{c}{p-value} &
  \multicolumn{1}{c}{$beta_6$} &
  p-value \\ \hline
GPT-4-Turbo Judge &
  -0.463 &
  \textless{}.001 &
  -0.139 &
  \multicolumn{1}{c}{0.138} &
  -0.410 &
  \textless{}.001 &
  0.402 &
  \textless{}.001 &
  0.023 &
  0.790 \\
GPT-3.5-Turbo Judge &
  0.369 &
  \textless{}.001 &
  -0.478 &
  \textless{}.001 &
  -0.222 &
  0.008 &
  -0.334 &
  0.001 &
  0.563 &
  \multicolumn{1}{l}{\textless{}.001} \\
Gemini-Pro Judge &
  -0.005 &
  0.955 &
  -0.812 &
  \textless{}.001 &
  -0.531 &
  \textless{}.001 &
  -0.683 &
  \textless{}.001 &
  0.103 &
  0.290 \\ \hline
\end{tabular}
\end{table}
\section{Coefficients of SGC conditions}
\label{appx:SGC conditions}
The logistic regression and Wald test results of coefficients of SGC conditions are shown in Table~\ref{tab:irre-logistic-regression}.

The coefficient are not quite suitable for being used to analyse the self-enhancement bias of LLM Judges, since these coefficients are strongly influenced by the confounding factor of notable position bias of LLM Judges, and there is no coefficient for the 6) SGC condition as mentioned in paper Sec. 3.4.3.

\newpage

\section{\chadded{Human expert labelling rubric}}
\chadded{
The following coding book is first discussed and developed among 2 researchers in self-regulated learning and educational technology, based on the theory and definition of the process library of COPES model (see Appendix B), and then refined by them in two rounds of inter-rater tests.
}
\subsection{\chadded{Supported SRL processes in scaffolds}}

\setlength{\LTleft}{0pt}
\setlength{\LTright}{0pt}
\captionsetup[longtable]{labelformat=empty, justification=raggedright}
\begin{longtable}{p{0.10\textwidth}p{0.22\textwidth}p{0.30\textwidth}p{0.36\textwidth}}
\caption{\parbox{\linewidth}{\raggedright\textbf{Table 12}\\Labelling Rubric}}
\label{tab:labelling_criteria}\\
\toprule
\textbf{SRL Processes} & \textbf{Constructs} & \textbf{Definition} & \textbf{Assignment Criteria} \\
\midrule
\endfirsthead
\caption[]{\parbox{\linewidth}{\raggedright\textbf{Table 12 (continued)}}} \\
\toprule
\textbf{SRL Processes} & \textbf{Constructs} & \textbf{Definition} & \textbf{Assignment Criteria} \\
\midrule
\endhead
\midrule
\multicolumn{4}{r}{{Continued on next page}} \\
\midrule
\endfoot

\bottomrule
\endlastfoot
      O.M.2 &                                                        Create\_Highlight &                                                                                  Learners create highlights on the reading materials &                                                                                     Assign this when scaffolds mention highlight or highlighter or similar meanings like 'draw upon the material'. 'Mention' means anywhere in the scaffold. \\
    C.SAR.1 &                                                        Table\_Of\_Content &                                                                                                  Learners check the table of content &                                                             Assign this when scaffolds mention table of contents or the Contents in any way, or 'gain an overview of your learning material' or 'see how each section is related to others'. \\
    C.SAR.2 &                                                           Try\_Out\_Tools &                                   Learners quickly (less than 3 seconds) open and close tools for the first time without using them. &                                                                                                                                                                   Assign this when scaffolds ask learners to try for tools or mention tools. \\
    C.STR.2 &  Task\_Overview / Task\_Requirement / Learning\_Goal / Rubric (first time) & For the first time that learners open task overview/requirement or learning goal or rubric page to read about what the task is about &                Assign this when scaffolds mention task overview or requirements or rubric or learning goals in any way, and do not specify if it's the second time or after the first time (e.g. 'remind you' implies it's the second time). \\
      O.S.3 &                                                         Page\_Navigation &                                                                   Learners navigate through several pages (stay less than 6 seconds) &                                                                                                                                         Assign this when scaffolds mention navigating or reviewing pages or paging or navigating in any way. \\
  S.ASBTS.2 &                                                            Open\_Planner &                                                                   Learners open the planner tool and read or think about their plans &                                                                                                                                                                                          Assign this when scaffolds mention planner or plan. \\
      O.M.3 &                                     Read\_Annotation / Delete\_Annotation &                          Learners click on or delete their annotations or open the annotation tool to read their notes or highlights & Assign this when scaffolds mention the meaning of reading or deleting notes or annotation or highlights or using the annotation tool in any way, but not read notes while reading materials because this is not to open the annotation tool. \\
    C.MTC.1 &                                                                   Timer &                                                                              Learners click and check the time left using timer tool &                                                                                                                                                       Assign this when scaffolds mention timer or the meaning of looking at time in any way. \\
    C.MTR.2 & Task\_Overview / Task\_Requirement / Learning\_Goal (after the first time) &              Again that learners open task overview/requirement or learning goal or rubric page to read about what the task is about &                                                                                             Assign this when scaffolds mention task overview or requirements or rubric or learning goals in any way, and do not specify it's the first time. \\
      O.A.3 &                                                         Pastetext\_Essay &                                                           Learners copy and paste materials from reading content to the essay window &                                                                     Assign this when scaffolds mention the meaning or words of copying or pasting or directly mentioning 'O.A.3' with explanation, except 'do not just paste' and 'quoting'. \\
      O.R.2 &                                                  Write\_Essay\_Rehearsing &                                                                 Learners open the essay and write to rehearse materials from reading &                                                                                              Assign this when scaffolds mention including materials or contents or annotation or information into the essay, or directly mention rehearsing. \\
      O.T.2 &                                                 Write\_Essay\_Translating &                                                                Learners open the essay and write to translate materials from reading &          Whenever 'translate' or 'translation' are mentioned in any way; When adding novel information to the essay is mentioned; When related meanings referring to writing essays according to materials are mentioned but not rehearsing. \\
      O.S.1 &                                                       Search\_Annotation &                                                            Learners use search annotation tool to search and check their annotations &                                                                                                                                                                                     Assign this when scaffolds mention searching in any way. \\
      O.T.1 &                                                             Create\_Note &                                                                                                      Learners create notes and write &                                                                                                                                                                                Assign this when scaffolds mention writing or creating notes. \\
      O.M.1 &                                                        Label\_Annotation &                                            Learners label or add new labels, or accept suggested labels on their notes or highlights &                                        Assign this when scaffolds mention the annotation tool or label annotation or label or similarly like adding bookmark or creating annotation, but not 'check the annotation' or 'reviewing bookmark'. \\
\end{longtable}

\subsection{\chadded{Hallucinations in scaffolds}}
\chadded{
To label apparent hallucinations that violate scaffolding instructions in the prompts, for each scaffold, label results and explanations of three GPT-4-Turbo API call results were provided to coders. Additionally, coders were especially aware of the following four kinds of violations of scaffolding instructions in the prompts, as concluded in inter-rater tests.
}
\begin{enumerate}
    \item \textbf{Directly mention 'O.T.2' the names:} \\
    Break the scaffold rules in the prompt. Examples: 'I am halfway through my essay about AI in medicine and I have about 15 minutes left to complete it.', 'in this prompt'; it did not understand it's providing scaffold.
    
    \item \textbf{Understand O.T.2 as text translation:} \\
    Examples: 'using text translation tool'; '100-200 words' 'in the past 15 minutes'; it did not understand SRL process meanings or misunderstanding task description given in the prompt.
    
    \item \textbf{Writing suggestions misunderstanding:} \\
    Examples: 'Please write in 300-400 words' 'writing more about how certain AI applications could revolutionize healthcare will give your essay depth and detail.', it did not understand you're not giving suggestions on how to write.
    
    \item \textbf{Fail to give feedback in only one paragraph as required in the prompt instruction to generate the scaffolds.}
\end{enumerate}

% \clearpage
%TC:endignore
\printcredits
\bibliographystyle{cas-model2-names}
\bibliography{cas-refs}

\end{document}